\documentclass{agujournal2019}
\usepackage{url} 
\usepackage[inline]{trackchanges} 
\usepackage{soul}
\usepackage[utf8]{inputenc} 
\usepackage[T1]{fontenc}    
\usepackage{hyperref}
\usepackage{url}            % simple URL typesetting
\usepackage{booktabs}       % professional-quality tables
\usepackage{amsfonts}       % blackboard math symbols
\usepackage{nicefrac}       % compact symbols for 1/2, etc.
\usepackage{microtype}      % microtypography
\usepackage{lipsum}
\usepackage{graphicx}
\usepackage{natbib}
\usepackage{doi}
\graphicspath{ {./images/} }
\usepackage{amssymb}
\usepackage{latexsym}
\usepackage{amsmath} 
\usepackage{epsfig}
\usepackage{pdfpages}
\usepackage{graphicx}
\usepackage{subcaption}
\usepackage{mathtools}
\usepackage{setspace}
\usepackage{academicons}
\usepackage{caption} 
\newcommand{\vectorproj}[2][]{\textit{proj}_{\vect{#1}}\vect{#2}}
\newcommand{\vect}{\mathbf}
\usepackage{orcidlink}
\usepackage[section]{placeins}
\usepackage{nicematrix}

\journalname{Space Weather}

\begin{document}

%%%%%%%%%%%%%%%%%%%%%%%%%%%%%%%%%%%%%%%%%%%%%%%
% TITLE

% (A title should be specific, informative, and brief. Use
% abbreviations only if they are defined in the abstract. Titles that
% start with general keywords then specific terms are optimized in
% searches)

%%%%%%%%%%%%%%%%%%%%%%%%%%%%%%%%%%%%%%%%%%%%%%%

\title{Visibility Analysis of the Sun as Viewed from Multiple Spacecraft at the Sun-Earth Lagrange Points}

\authors{Jinsung Lee\affil{1} \orcidlink{0000-0001-9671-067X},
Sung-Hong Park\affil{2} \orcidlink{0000-0001-9149-6547},
Arik Posner\affil{3} \orcidlink{0000-0003-1572-8734},
Kyung-Suk Cho\affil{2} \orcidlink{0000-0003-2161-9606},
Jaemyung Ahn\affil{1} \orcidlink{0000-0003-4971-5130}}

\affiliation{1}{Korea Advanced Institute of Science and Technology, Department of Aerospace Engineering, 291 Daehak-ro, Daejeon 34131, Republic of Korea}
\affiliation{2}{Korea Astronomy and Space Science Institute, 776 Daedeok-daero, Daejeon 34055, Republic of Korea}
\affiliation{3}{NASA/HQ, Washington, DC, USA}

% Example: \correspondingauthor{First and Last Name}{email@address.edu}

\correspondingauthor{Sung-Hong Park}{shpark@kasi.re.kr}

\begin{keypoints}
\justifying
\item Multi-spacecraft observations of the Sun from the Sun-Earth Lagrange points L1, L4 and L5 can provide a clear and wide-angle view to investigate the Sun-Earth, Sun-Moon and Sun-Mars connections
\item Visibility of the solar surface is analyzed based on remote-sensing observations from single (L1), double (L1 and L4) and multi-space missions (L1, L4 and L5)
\item A quantitative comparison of the solar surface visibility is made in the context of (1) observation days per year for a given solar latitude, (2) a chance of observing a limb flare from one spacecraft and its on-disk counterpart from the others, and (3) continuous tracking of a target feature (such as sunspots) on the solar disk
\end{keypoints}

\begin{abstract}
\justifying
Beyond the Sun-Earth line, spacecraft equipped with various solar telescopes are intended to be deployed at several different vantage points in the heliosphere to carry out coordinated, multi-view observations of the Sun and its dynamic activities. In this context, we investigate solar visibility by imaging instruments onboard the spacecraft orbiting the Sun-Earth Lagrange points L1, L4 and L5, respectively. An optimal arrival time for vertical periodic orbits stationed at L4 and L5 is determined based on geometric considerations that ensure maximum visibility of solar poles or higher latitudes per year. For a different set of orbits around the three Lagrange points (L1, L4 and L5), we calculate the visibility of the solar surface (i.e., observation days per year) as a function of the solar latitude. We also analyze where the solar limb viewed from one of the three Sun-Earth Lagrange points under consideration is projected onto the solar surface visible to the other two. This analysis particularly aims at determining the feasibility of studying solar eruptions, such as flares and coronal mass ejections, with coordinated observations of off-limb erupting coronal structures and their on-disk magnetic footpoints. In addition, visibility analysis of a feature (such as sunspots) on the solar surface is made for multiple spacecraft in various types of orbits with different inclinations to quantify the improvement in continuous tracking of the target feature for studying its long-term evolution from emergence, growth and to decay. A comprehensive comparison of observations from single (L1), double (L1 and L4) and multi-space missions (L1, L4 and L5) is carried out through our solar visibility analysis, and this may help us to design future space missions of constructing multiple solar observatories at the Sun-Earth Lagrange points.
\end{abstract}

\section*{Plain Language Summary}
\justifying
The Heliophysics community and space weather service providers demand for their research and model operation to send more spacecraft equipped with various scientific payloads beyond the line connecting the Sun and Earth. These spacecraft will be strategically positioned to observe the Sun and its activities from multiple perspectives. We specifically focus on studying the visibility of the Sun from spacecraft located at the Sun-Earth Lagrange points L1, L4 and L5, where the gravity of the Earth and Sun allow the spacecraft to remain stable over long periods of time. We determined an optimal time for spacecraft to enter orbits around L4 and L5 based on the specific purpose of maximizing the visibility of the solar poles and higher latitudes throughout the year. We also examine whether semi-circular, loop-shaped structures near the solar limb viewed from one of the Lagrange points are visible to the other Lagrange points. This analysis helps us to estimate the feasibility condition of studying solar eruptions (such as flares and coronal mass ejections) in the context of their coordinate observations: i.e., erupting coronal structures above the solar limb from one spacecraft and their source region on the solar disk from the others. Furthermore, we assess the visibility of specific features, such as sunspots, on the solar surface for multiple spacecraft in different types of orbits with varying inclinations, which enables us to quantify the improvement of continuous tracking of these features for the study of their long-term evolution from emergence and growth to decay.

\section{Introduction\label{sec:intro}}
The Sun consistently blows out the high-speed solar wind into interplanetary space and occasionally produces large eruptive events such as flares, solar energetic particles (SEPs) and coronal mass ejections (CMEs). When such dynamic solar activities occur, they may cause heliospheric disturbances and consequently affect the so-called space weather \cite[e.g.,][]{schwenn2006,vourlidas2021,temmer2021,zhang2021}. The magnetic field of the Sun is fundamental to understanding the origin, structure and dynamics of the solar wind \cite[e.g.,][]{wang2000,antiochos2011} as well as the trigger mechanism of solar eruptions \cite[e.g.,][]{park2010,kusano2012}. It is therefore important to acquire information on the dynamic solar activity and magnetic field extended to the heliosphere with remote-sensing observations as precisely, immediately and extensively as possible. There have been, however, limitations to obtaining such information due to the observations mostly carried out from the Sun-Earth line (i.e., from ground telescopes, Earth-orbiting satellites, or spacecraft at the Sun-Earth Lagrange point L1). To overcome this limited view of the Sun, various space missions are currently in operation (e.g., Parker Solar Probe, Solar Orbiter, STEREO-A) and in preparation (e.g., ESA’s Vigil at the Sun-Earth Lagrange point L5) off the Sun-Earth line, thereby carrying out coordinated, multi-view observations of the Sun. 

\begin{figure}[htbp]
  \centering
  \includegraphics[page=1, scale=0.6]{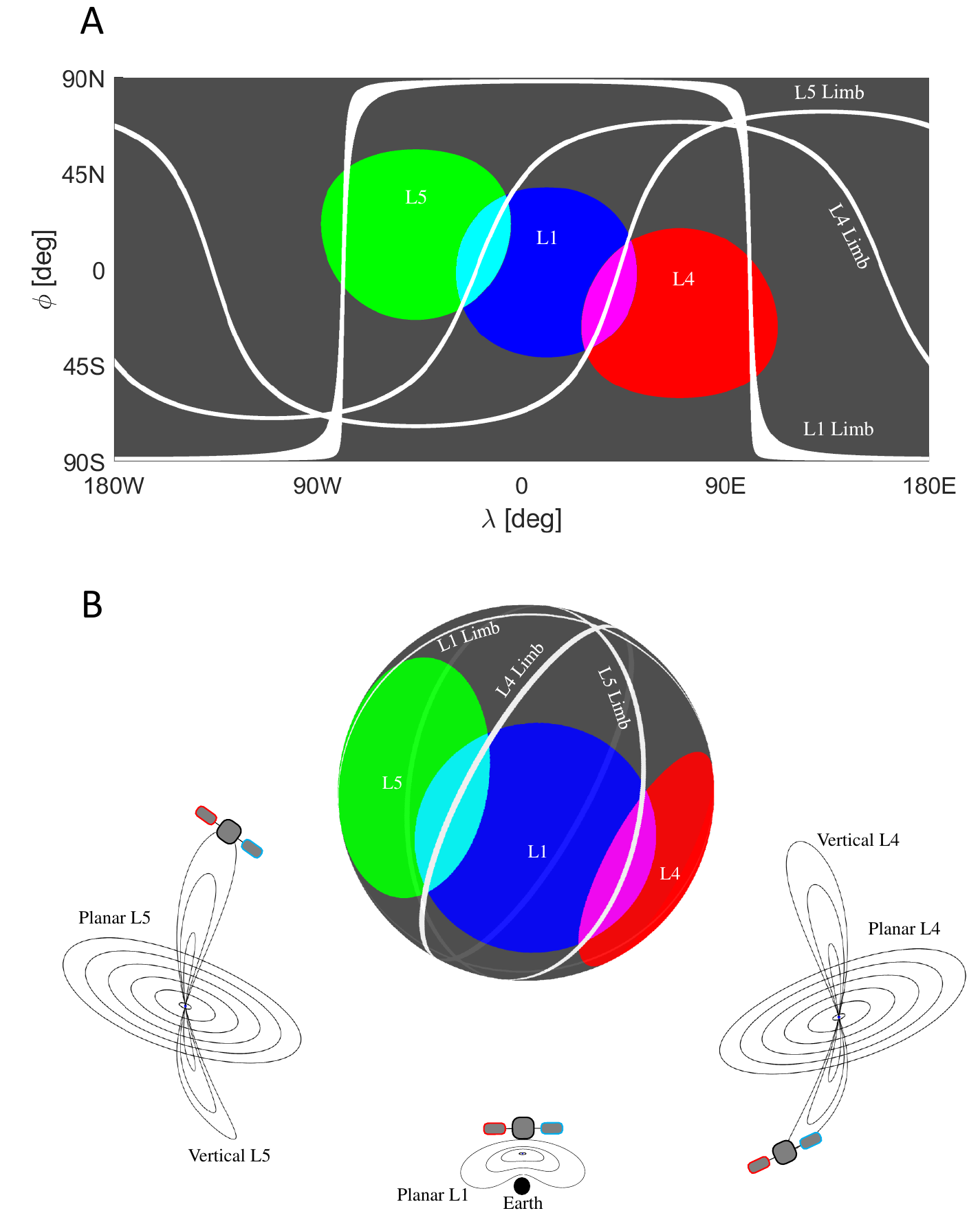}
  \caption{Cartoon showing solar surface visibility with three spacecraft positioned at the Sun-Earth Lagrange points L1, L4 and L5 on the Stonyhurst heliographic coordinate system (panel a) and on the spherical surface of the Sun (panel b). The visible solar surface by an imaging instrument (with a viewing angle of $40^\circ$ toward the center of the Sun) onboard each of the spacecraft is shown in blue, red and green, respectively, for L1, L4 and L5, while the invisible region in gray. The region visible by two spacecraft is denoted in magenta for L1 and L4, while in cyan for L1 and L5. The white line represents the solar limb viewed from each of the spacecraft.}
  \label{fig:300degcoverage}
\end{figure}

\begin{figure}[htbp]
  \centering
  \includegraphics[page=2, scale=0.85]{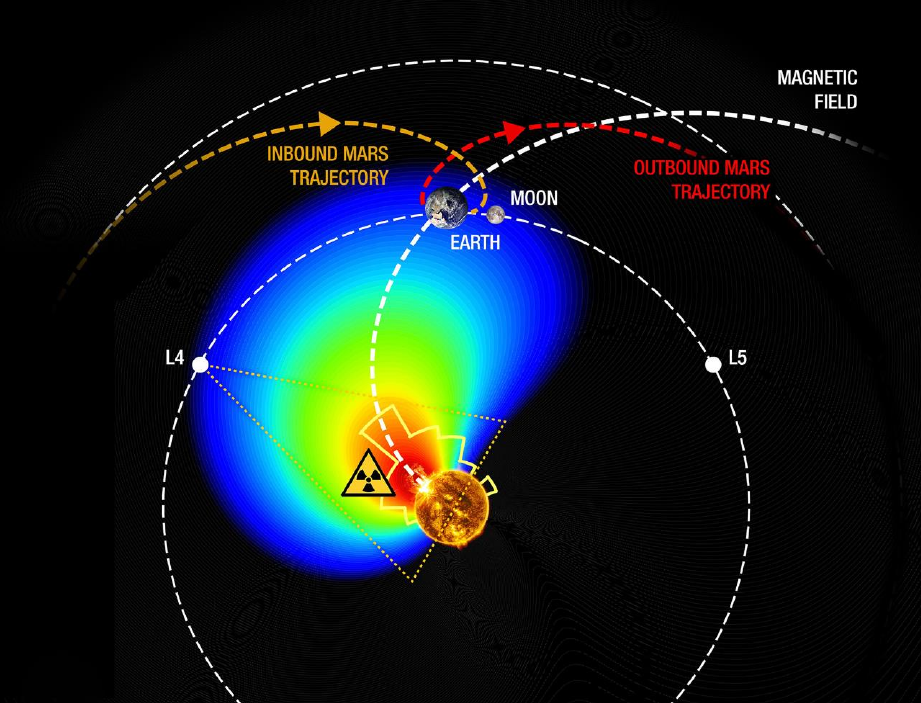}
  \caption{Illustration of the solar radiation hemisphere that is the relative solar hemisphere from a 1 AU observer at the Earth with possessing the potential to severely affect the radiation environment at or near the Earth. It ranges from 30°E to 150°W in solar longitudes and is best observed from L4. A histogram from \cite{richardson201425} displays source longitudes of major solar proton events in the energy range of 14–-24 MeV. Highlighted are the Earth-Moon system, Hohmann transfer orbits to and from Mars, L4 and L5, and an interplanetary magnetic field line from 60°W on the Sun to the Earth modeled from an average solar wind speed in the ecliptic plane. This figure is adapted from \cite{posner2021multi}.}
  \label{fig:posnerFig}
\end{figure}

The Sun-Earth Lagrange points (hereafter, simply referred to as L1--L5), especially the triangular points L4 and L5, are viable options for observing the Sun from multiple angles \citep{cho2023opening,moon2024}. Observations made through remote sensing, spanning heliographic longitudes up to $120^\circ$ apart between L4 and L5, provide visibility coverage for a substantial portion (i.e., $300^\circ$ of longitude as the maximum meridional extent allowed) of the solar surface at any given time. This effectively reduces any observational blind spots on the solar surface. Such wide-view coverage is pivotal for real-time monitoring of the Sun, improving the capability to detect and monitor solar eruptive events and predict their impact on the heliospheric environment. As shown in Figure~\ref{fig:posnerFig}, remote-sensing observations from L4, together with L1, can help to carry out a thorough investigation of solar source regions that produce SEPs. The wider field of view for solar observations also enables the longer tracking of a feature of interest, which will help us to understand its persistent evolution as it rotates: e.g., emergence, development and decay of sunspots \cite[e.g.,][]{driel2015}.   

The strategic positioning of spacecraft at the multiple Sun-Earth Lagrange points will provide an advantage in reconstructing the three-dimensional (3D) structure of CMEs and determining their propagating direction and speed from stereoscopic observations. These multiple viewpoints can also contribute to resolving an intrinsic $180^\circ$ ambiguity in the orientation of the transverse magnetic field present in Zeeman effect measurements on the surface, as discussed in \citet{posner2021multi,valori2023stereoscopic,cho2023opening,moon2024}. In addition, a combination of solar on-disk and off-limb observations for the same target region by this multi-spacecraft configuration will be beneficial for a better understanding of solar eruptions, allowing us more comprehensive modeling of their magnetic field structure and evolution in the 3D solar atmosphere (i.e., through multiple layers from the photosphere to the corona). 

A unique perspective to precisely observe the Sun's high latitudes can be achieved by optimizing the design of vertical periodic orbits for spacecraft at L4 and L5, reducing the projection effect on solar image data. The polar and high-latitude region of the Sun is of particular interest because the magnetic field strength in such region around the minimum of the 11-year solar activity cycle has a close relation with the sunspot number at the upcoming maximum \cite[e.g.,][]{schatten1978,pesnell2012,javaraiah2023}. One can accurately measure solar meridional flows at high latitudes (toward the polar region), which are thought to play an important role in weakening the Sun’s polar magnetic fields and eventually changing their polarity \cite[e.g.,][]{hathaway2010,nandy2011,lekshmi2019}.

This paper is structured in the following manner. In Section~\ref{sec:orbits}, we first introduce periodic orbits associated with the Sun-Earth system's first (L1), fourth (L4) and fifth (L5) Lagrange points based on the dynamics of the Circular-Restricted Three-Body Problem (CRTBP). Section~\ref{sec:methods} presents our methodology for analyzing solar visibility. Section~\ref{sec:results} discusses the visibility analysis results of the Sun as viewed from a single spacecraft at L1 as well as from multiple spacecraft at L1, L4 and L5. Section~\ref{sec:summary} summarizes the study and discussions for future studies.

\section{Orbits around the Sun-Earth Lagrange points\label{sec:orbits}}
%This section addresses the issue of determining the solar surface visibility from multiple spacecraft placed in the Sun-Earth periodic orbit. First, we introduce the circular restricted three-body problem dynamics. Following that, we delve into the discussion of periodic orbits, particularly those revolving around the triangular (L4 and L5) and the first Lagrange points (L1). Lastly, we explore the criteria for determining the periodic orbits for solar-surface and limb observations.

\subsection{Circular Restricted Three-body Problem\label{subsec:orbits_crtbp}}
The Circular Restricted Three-Body Problem (CRTBP) governs the motion of a massless spacecraft influenced by the gravitational attraction of two large masses. In the case of the Sun-Earth system, the origin of the CRTBP rotating frame is located at the barycenter of the two masses. The x-axis consistently points from the system's primary to the secondary masses (i.e., from the Sun to the Earth). The z-axis aligns with the secondary body's angular momentum vector. Employing canonical units in Table~\ref{tab:nond} for dimensionless quantities as well as setting the angular velocity of the primary and secondary with respect to the barycenter as unity, the governing CRTBP equations are defined with the position vector $\mathbf{r}=\left[ x,y,z\right]$ and the velocity vector $\mathbf{v}=\left[ v_x,v_y,v_z\right]$, as follows:
\begin{equation}
    \dot{\mathbf{x}}=\mathbf{f}\left( \mathbf{x} \right)=\left[ v_x,v_y,v_z,2v_y+\frac{\partial\Omega}{\partial x},-2v_x+\frac{\partial\Omega}{\partial y} ,\frac{\partial\Omega}{\partial z}\right].
\end{equation}
The potential function $\Omega$ is defined as:
\begin{equation}
    \Omega\left(\mathbf{r}\right) = \frac{x^2+y^2}{2}+\frac{1-\mu}{r_1}+\frac{\mu}{r_2}
\end{equation}
where $\mu=\frac{m_2}{m_1+m_2}$, and the positions of the spacecraft measured from the primary and secondary masses are given by $r_1=\sqrt{\left( x+\mu \right)^2+y^2+z^2 }$ and $r_2=\sqrt{\left( x-1+\mu \right)^2+y^2+z^2 }$, respectively \citep{vallado2001fundamentals}. The CRTBP dynamical system features a conserved integral known as the Jacobi integral $\left(C\right)$:
\begin{equation}\label{jacobi}
    C=2\Omega-\left(v_x^2+v_y^2+v_z^2\right).
\end{equation}
The Jacobi integral $C$ determines the so-called Hill's region, which, in turn, defines the allowable region of motion in the rotating frame for a given energy level. More on the Jacobi's integral on the trajectories near the periodic orbits can be found in \citet{lee2021successive,lee2022poincare,lee2024low,lee2023manifold}.

\begin{table}
    \centering
    \caption{Non-dimensional canonical unit of the Sun-Earth CRTBP system}
    \begin{tabular}{cc}
        \hline
        Unit & Value\\
        \hline
        Time (TU)& 5.0226757E6 seconds \\
        Distance (AU) & 1.4959965E8 km\\
        Velocity (VU)& 29.7849 km/s\\
        \hline
    \end{tabular}
    \label{tab:nond}
\end{table}

\subsection{Periodic Orbits of the Sun-Earth Lagrange Points\label{subsec:orbits_lagrange}}
\begin{figure}[htbp]
  \centering
  \includegraphics[page=3, scale=0.7]{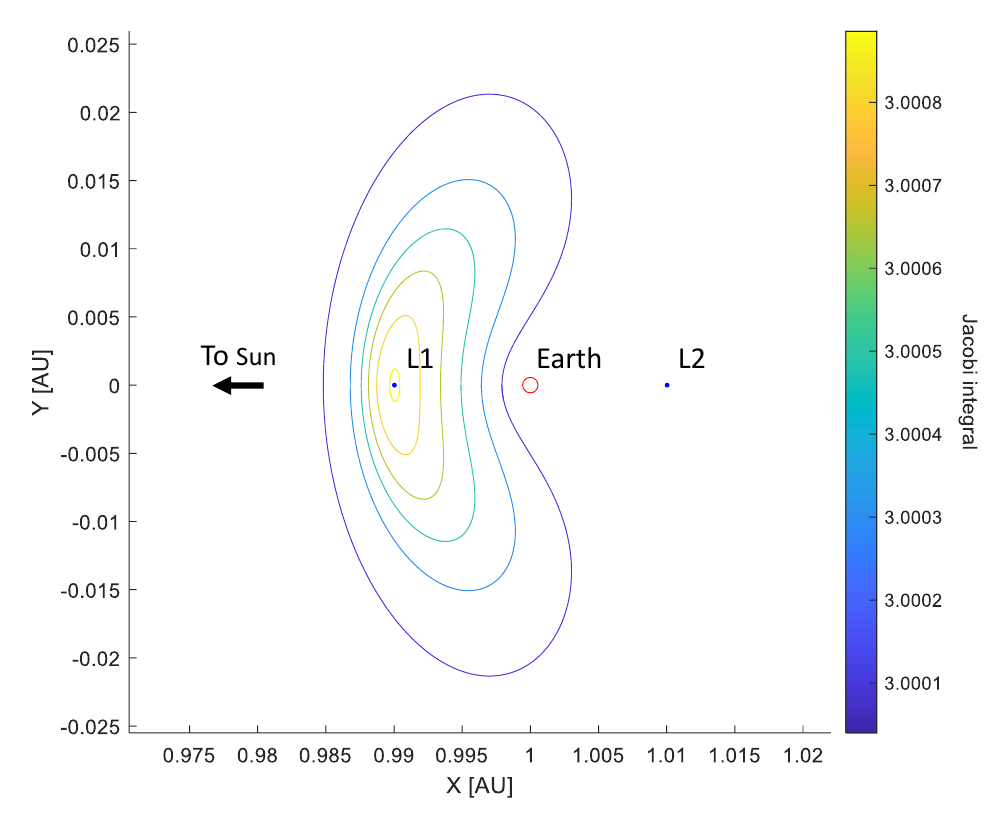}
  \caption{Planar Lyapunov periodic orbits at the Sun-Earth Lagrange point L1 are shown with a different set of values for the Jacobi integral.}
  \label{fig:L1PO}
\end{figure}

\begin{figure}[htbp]
  \centering
  \includegraphics[page=4,scale=0.7]{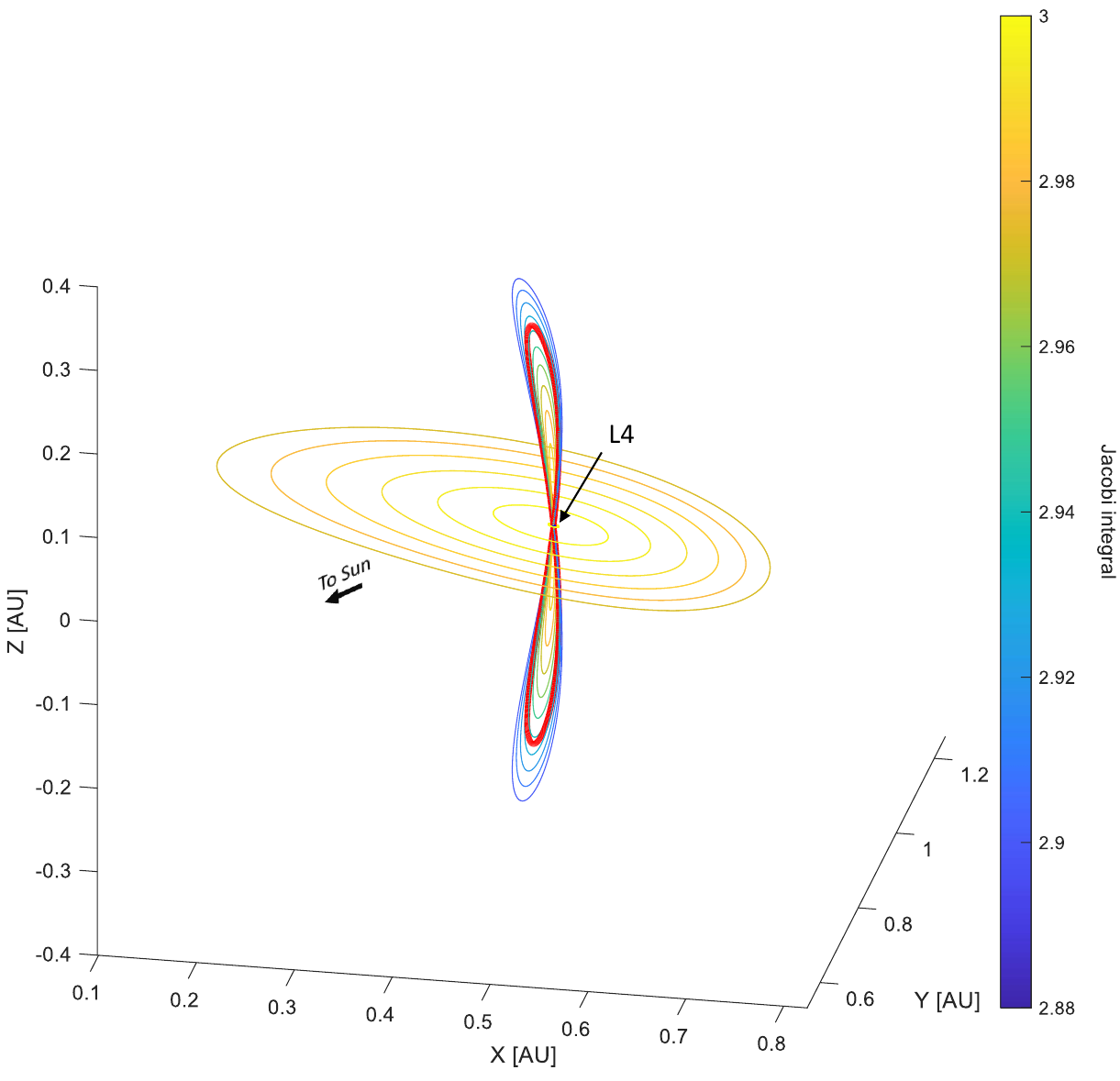}
  \caption{Planar and vertical Lyapunov periodic orbits at L4 are shown with a different set of values for the Jacobi integral. For the vertical periodic orbits, the one with an inclination of $14.5^\circ$ with respect to the ecliptic plane is marked in red.}
  \label{fig:L4PO}
\end{figure}

In the context of CRTBP, periodic orbits appear as a distinctive trajectory circumnavigating the Lagrange points. When represented within the rotating frame, these orbits manifest as recurrent, ensuring that the spacecraft's position and velocity return to their initial conditions. The stability of these periodic orbits is intrinsically linked to that of the corresponding Lagrange points. Periodic orbits around L1, L2 and L3 are unstable, while those around L4 and L5 are stable \citep{vallado2001fundamentals}. An infinite spectrum of periodic orbits exists, each distinguished by its unique Jacobi integral. Among multiple families of periodic orbits, planar Lyapunov periodic orbits around L1 are presented in Figure~\ref{fig:L1PO}, which are symmetric with respect to the x-axis \citep{thurman1996geometry}. Meanwhile, Figure~\ref{fig:L4PO} shows planar and vertical Lyapunov periodic orbits around L4 at different energy levels. As the Jacobi integral decreases, both L1 and L4 Lyapunov orbits increase amplitudes in the x- and y-directions. Similarly, the lower value of the Jacobi integral corresponds to the larger amplitude in the z-direction for L4’s vertical periodic orbits. We note that these characteristics of L4’s periodic orbits remain the same for L5.

As suggested by its name, the planar Lyapunov orbit family has a consistent zero inclination across its orbits (i.e., lying in the x-y plane) with respect to the ecliptic plane. At the same time, there is a noticeable increase in eccentricity and a decrease in the semi-major axis as the Jacobi integral decreases. On the other hand, vertical periodic orbits, passing through the triangular Lagrange point (L4 or L5) twice a year, maintain an osculating semi-major axis of 1 AU and zero eccentricity. The orbital period of the periodic orbit varies slightly depending on the $C$ level. Nevertheless, it's crucial to highlight that the orbital period plays a less significant role for spacecraft orbiting around the Lagrange points as they synchronize with the Earth. Out of numerous periodic orbits, this study specifically focuses on (1) the planar Lyapunov periodic orbit around L1 with $3.0001 \le C \le 3.001$, (2) the planar Lyapunov periodic orbit around L4 (or L5) with $2.96 \le C \le 2.999$, and (3) the vertical periodic orbits around L4 (or L5) with $2.88 \le C \le 2.999$.

\section{Methods for Solar Visibility Analysis\label{sec:methods}}
\subsection{Definition of Solar Visibility Conditions\label{subsec:methods_definition}}
\begin{figure}[htbp]
  \centering
  \includegraphics[page=5, scale=0.4]{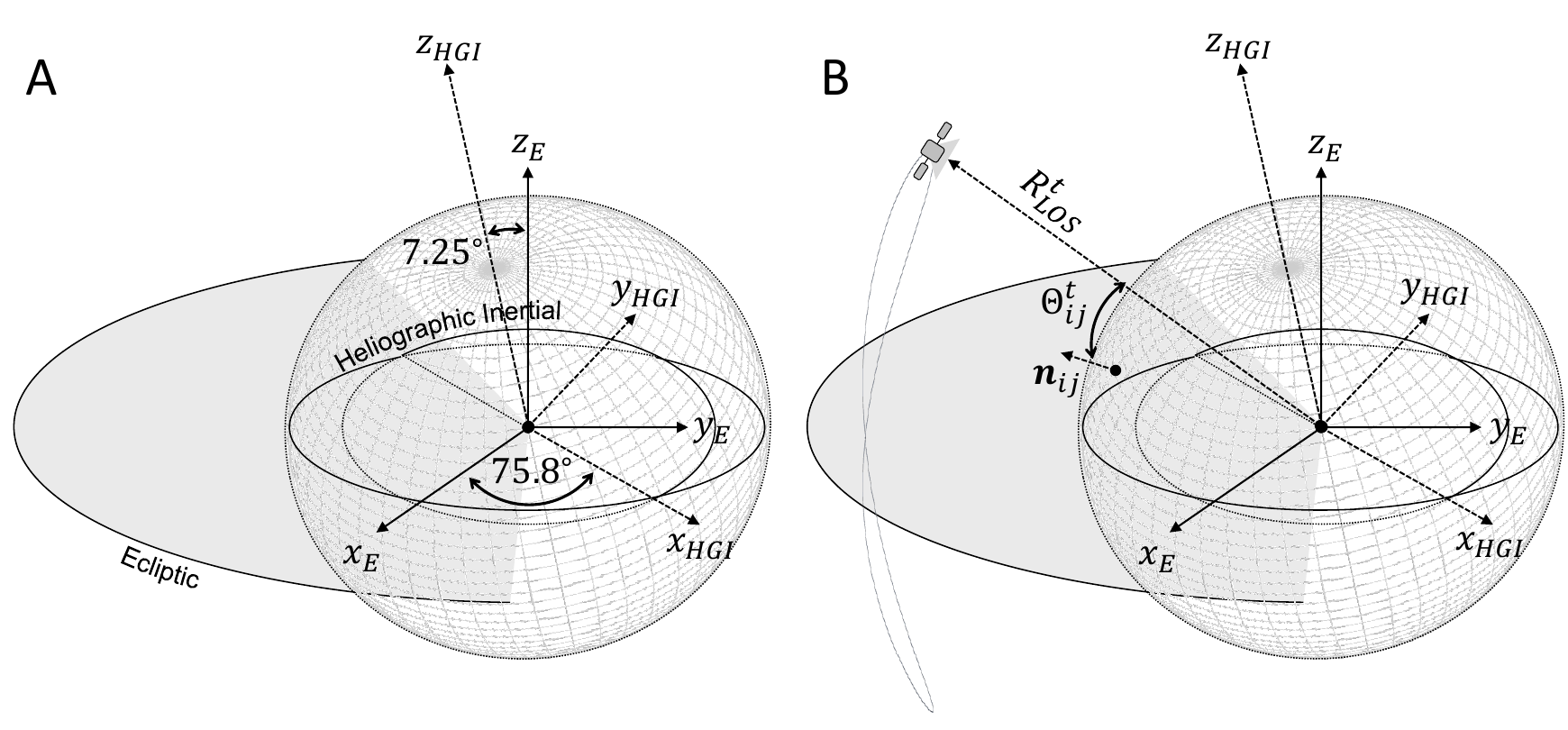}
  \caption{Definition of the heliographic inertial coordinate system (HGI) with respect to the ecliptic coordinate system (E). In panel (b), $\Theta^{t}_{ij}$ is an angle between the surface normal vector $\mathbf{n}_{ij}$ and the line-of-sight vector $\mathbf{R}_{LOS}^{t}$ at time $t$ of observations by an instrument onboard a spacecraft.}  
  \label{fig:HGI}
  \centering
\end{figure}

In this analysis of solar surface visibility, the heliographic inertial frame (HGI) serves as the appropriate coordinate system for determining the latitude and longitude coverage of the solar surface, as depicted in Figure~\ref{fig:HGI}{a}. The x-y plane of the HGI frame is defined as the solar equatorial plane, and the z-axis (i.e., the solar rotation axis) is inclined at $7.25^\circ$ relative to the z-axis of the ecliptic coordinate system. The HGI's x-axis is directed along the intersection line of the ecliptic and solar equatorial planes \citep{thompson2006coordinate}, where the ecliptic longitude of the Sun’s ascending node is $75.77^\circ$ (J2000).

This study incorporates the corrected oblateness of the non-magnetic Sun, quantified as $8.01\pm0.14$ milli-arcseconds or equivalently $2.225 \times 10^{-6}$ degrees \citep{dicke1974oblateness,rozelot2001theory}. The variance of $\pm0.14$ milli-arcseconds is omitted in the visibility analysis. The solar surface represented by the spherical grid in Figure~\ref{fig:HGI} is segmented into the grid cells of which each has longitudinal and latitudinal extents of $5^\circ$ and $6^\circ$, respectively. In Section~\ref{subsec:results_sunspot}, when analyzing the sunspot visibility, the differential rotational rate $\dot{\omega}\left(\phi\right)$ of the Sun is modeled as follows:
\begin{equation}
\dot{\omega}\left( \phi \right) = A + B\sin^2(\phi) + C\sin^4(\phi),
\label{differentialrotationEQ}
\end{equation}
where $A=14.713 \pm$ 0.0491$^\circ$/day, $B= -2.396 \pm$ 0.188$^\circ$/day, $C= -1.787 \pm$ 0.253$^\circ$/day, and $\phi$ denotes the heliographic latitude (positive/negative towards the North/South) measured from the solar equator \citep{snodgrass1990rotation}.

\begin{figure}[htb!]
\centering
  \begin{tabular}{@{}cccc@{}}
    \begin{subfigure}[b]{.80\linewidth}
        \centering
        \includegraphics[page=1,width=.99\textwidth]{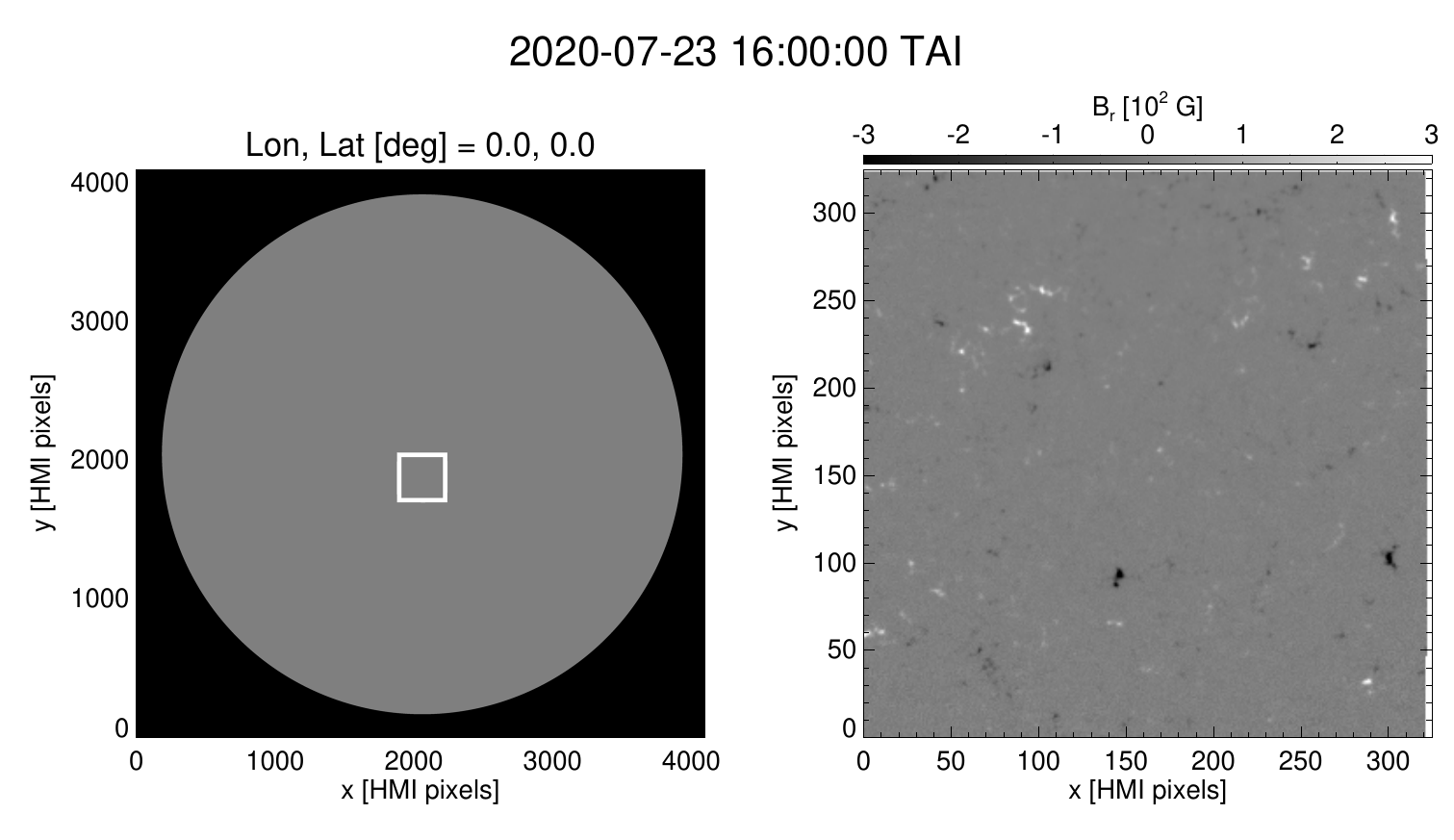}
        \caption{$\Theta=0$}
    \end{subfigure}\\%  
    \begin{subfigure}[b]{.80\linewidth}
        \centering
        \includegraphics[page=2,width=.99\textwidth]{Figures/ProjectionEffect.pdf}
        \caption{$\Theta=60$}
    \end{subfigure}
  \end{tabular}
  \caption{Demonstration of how the projection effect degrades a solar image obtained by remote-sensing observations. In panel (a), (right) an image of the radial component $B_{r}$ of the Sun's photospheric magnetic field obtained by SDO/HMI is shown, and (left) its field-of-view is marked with the white box in the full solar disk. We note that $\Theta^{t}_{ij}$ is $0^\circ$ at the center of the cutout image. In panel (b), (right) the same cutout image of $B_{r}$ in panel (a) is remapped as simply locating the field-of-view to a higher latitude of $60^\circ$N but the same longitude.}
  \label{fig:projectioneffect2}
\end{figure}
The solar surface visibility condition is determined with the so-called position angle $\Theta^{t}_{ij}$ defined as:
\begin{equation}
\Theta^{t}_{ij} = \cos^{-1}
\left(
\frac{\mathbf{n}_{ij}\cdot\mathbf{R}_{LOS}^{t}}{|\mathbf{n}_{ij}||\mathbf{R}_{LOS}^{t}|}
\right),
\label{Eq.positionAngle}
\end{equation}
where $\mathbf{n}_{ij}$ is the normal vector of the grid cell on the surface with the $i$'th and $j$'th heliographic latitude and longitude, respectively. $\mathbf{R}_{LOS}^{t}$ is defined as the line-of-sight (LOS) vector of an imaging instrument onboard a spacecraft from the center of the Sun at $t$. As shown in figure \ref{fig:HGI}{b}, $\Theta^{t}_{ij}$ is an angle between $\mathbf{n}_{ij}$ and $\mathbf{R}_{LOS}^{t}$ at time $t$ of observations. It is assumed in this analysis that the given spacecraft at the Sun-Earth Lagrange points performs perfect attitude control to keep the LOS vector $\mathbf{R}_{LOS}^{t}$ persistently directed to the center of the Sun. For images obtained from remote-sensing observations, a feature located in ($i$, $j$) on the solar surface can be severely affected by the so-called projection effect depending on $\Theta^{t}_{ij}$. For example, the physical length of the feature on the solar surface are scaled by $1/\cos{\Theta^{t}_{ij}}$. In Figure~\ref{fig:projectioneffect2}, we demonstrate the projection effect using an image of the radial component $B_{r}$ of the Sun's photospheric magnetic field obtained by the Helioseismic and Magnetic Imager (HMI) onboard the Solar Dynamics Observatory (SDO). The right panel of Figure~\ref{fig:projectioneffect2}(a) shows a cutout image of $B_{r}$ at 16:00 TAI on 2020-07-23, of which the field-of-view is marked with the white box in the left panel. We note that $\Theta^{t}_{ij}$ is $0^\circ$ at the center of the cutout image. In Figure~\ref{fig:projectioneffect2}(b), the same cutout image in Figure~\ref{fig:projectioneffect2}(a) is remapped as simply locating the field-of-view to a higher latitude of $60^\circ$N but the same longitude. The image quality degrades as small-scale magnetic features on the solar surface become smoothed or their fine structures are unresolved due to the projection effect. Depending on science objectives and data requirements, it is required to investigate how critical the projection effect will impact images obtained from remote-sensing observations. In this context, given $\Theta^{t}_{ij}$, we determine the condition of the solar surface visibility (i.e., "good enough" to be used as scientific data or not), considering a different set of values for the maximum angle $\Theta_\texttt{max}$ allowed for on-disk observations as follows. 
\begin{equation}
\psi^{t}_{ij}=
\begin{cases}
    1& \text{if} \quad\Theta^{t}_{ij}<\Theta_\texttt{max}\\
    0& \text{otherwise}
\end{cases}
\label{ondiskeq}
\end{equation}
Here $\psi^{t}_{ij}$ is the element of the instantaneous visibility matrix at time $t$. For limb observations, both $\Theta_\texttt{max}$ and the minimum angle $\Theta_\texttt{min}$ are considered.
\begin{equation}
\psi^{t}_{ij}=
\begin{cases}
    1& \text{if} \quad\Theta_\texttt{min}<\Theta^{t}_{ij}<\Theta_\texttt{max} \\
    0& \text{otherwise}
\end{cases}
\label{limbeq}
\end{equation}

To determine the duration of the solar surface visibility per orbital period, $\psi^{t}_{ij}$ is then (1) multiplied with the simulation time step $\delta t$ of 2 hours, (2) summed over all simulation times of $T=\left\{ t_0,t_1,...,t_N \right\}$ during the period $\tau$ of each orbit considered at the Sun-Earth Lagrange points (i.e., $\tau=t_{N}-t_{0}=$ 1 year), and (3) divided by $\tau$:
\begin{equation}
    \Psi_{ij}=
    \frac{\sum\limits_{t \in T}\psi^{t}_{ij} \cdot \delta t}{\tau}.
\label{sumtimeeq}
\end{equation}
Here, $\Psi_{ij}$ represents the visibility duration of the heliographic coordinates ($i$, $j$) per orbital period with the unit of days per year. For a given latitude $i$, as calculating the average of $\Psi_{ij}$ values over all longitudes, we can determine the mean visibility duration $\bar{\Psi}_{i}$ as a function of latitude:
\begin{equation}
    \hat{\Psi}_{i}=
    \frac{\sum\limits_{j}\Psi_{ij}}{N_{j}},
\label{sumlongitudeeq}
\end{equation}
where $N_{j}$ is the total number of longitudinal grid points.

To analyze sunspot visibility, we define a single grid cell located at a heliographic latitude $\phi$ and longitude of $0^\circ$ in the simulation coordinate system at $t_0$ as a test sunspot. The test sunspot is set to rotate at a constant speed governed by the differential rotation described in Equation~\ref{differentialrotationEQ} and stays on the solar surface over the simulation period $\tau$. The normal vector $\mathbf{n}_\mathrm{spot}(t,\phi)$ of the test sunspot cell is then represented as follows:
\begin{equation}
    \mathbf{n}_\mathrm{spot}(t,\phi)=\mathcal{R}(\mathbf{z}_{HGI},\dot{\omega}(\phi)\cdot \Delta t)\mathbf{n}_\mathrm{spot}(t_s,\phi),
\end{equation}
where $\mathcal{R}(\mathbf{z}_{HGI},\dot{\omega}(\phi)\cdot \Delta t)$ is the rotation matrix of the sunspot over a time span of $\Delta t$ from $t_0$. The rotation matrix is defined as:
\begin{eqnarray}
    \mathcal{R}(\mathbf{z}_{HGI},\dot{\omega}(\phi)\cdot \Delta t) & = & \cos(\dot{\omega}(\phi)\cdot \Delta t)I_{3\times3} + \sin(\dot{\omega}(\phi)\cdot \Delta t)\mathbf{z}_{HGI}^\times \nonumber \\
    && + [1-\cos(\dot{\omega}(\phi)\cdot \Delta t)]\mathbf{z}_{HGI}\mathbf{z}_{HGI}^T,
\end{eqnarray}
where $\mathbf{z}_{HGI}^\times$ defines the skew matrix of $\mathbf{z}_{HGI}$. The mean visibility duration of the test sunspot per orbital period is determined with $\mathbf{n}_\mathrm{spot}(t,\phi)$, applying the same criterion (i.e., Equation~\ref{ondiskeq}) and the procedure (i.e., Equations~\ref{sumtimeeq} and~\ref{sumlongitudeeq}) as used in the analysis of the solar surface visibility.

\subsection{Setup for an Orientation of Vertical Periodic Orbits at L4 and L5~\label{subsec:methods_orientation}}

\begin{figure}[htbp]
  \centering
  \includegraphics[page=6, scale=0.45]{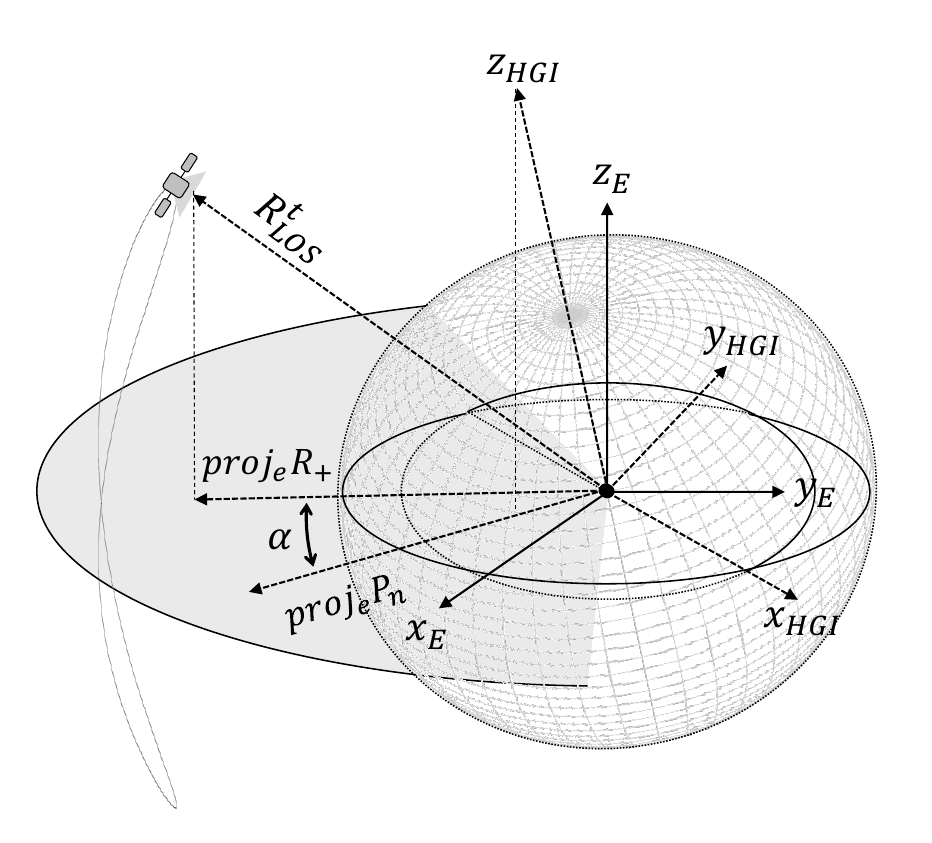}
  \caption{When a spacecraft is located at north of the ecliptic plane, $\alpha$ is defined as an angle between the two vectors $\vectorproj[E]{\mathbf{R_+}}$ and $\vectorproj[E]{P_N}$ projected onto the ecliptic frame. $\mathbf{R_+}$ is the LOS vector of an imaging instrument onboard a spacecraft from the center of the Sun when the spacecraft is located at the highest point on the z-axis of the ecliptic coordinate system, and $P_N$ is the vector directed from the solar center to the north pole.}
  \label{fig:alphadiagram}
  \centering
\end{figure}

\begin{figure}[htbp]
  \centering
  \includegraphics[page=7, scale=0.7]{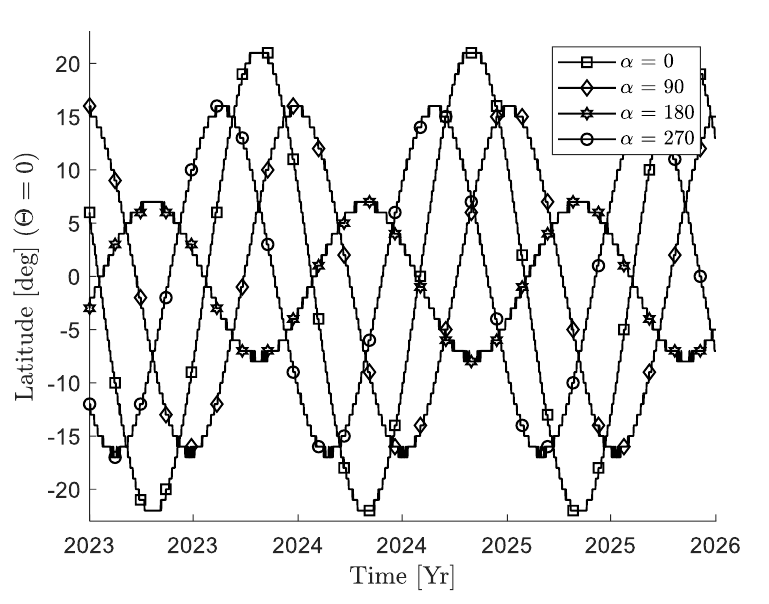}
  \caption{Time variation of the heliographic latitude for a solar surface grid cell with $\Theta^{t}_{ij}=0^\circ$ viewed from four spacecraft orbiting around L4 (or L5) with the same inclination of $14.5^\circ$ but different orientations (i.e., $\alpha=0^\circ$, 90$^\circ$, 180$^\circ$ and 270$^\circ$, respectively).}
  \label{fig:alpha_vs_lat}
  \centering
\end{figure}

\citet{posner2021multi} analyzed vertical periodic orbits at L4 and L5 to optimize the simultaneous view of high-latitude regions in both the northern and southern hemispheres of the Sun from remote-sensing observations at L1, L4 and L5 over a year. On the other hand, in this study, we aim to achieve a longer observation duration of high solar surface latitudes. In this context, we define $\alpha$ as $0^\circ$ as:
\begin{equation}
\alpha = 
\begin{cases}
  \measuredangle \left(\vectorproj[E]{\mathbf{R_+}} , \vectorproj[E]{P_N} \right) & \text{for spacecraft at north of the ecliptic plane,}\\
  \measuredangle \left(\vectorproj[E]{\mathbf{R_-}} , \vectorproj[E]{P_S} \right) & \text{for spacecraft at south of the ecliptic plane,}
\end{cases}
\label{Eq.Alpha}
\end{equation}
where $\vectorproj[E]{\left(\cdot\right)}$ denotes the vector projection onto the ecliptic plane, $\mathbf{R_{+/-}}$ is the LOS vector of an imaging instrument onboard a spacecraft from the center of the Sun when the spacecraft is located at the highest/lowest point on the z-axis of the ecliptic coordinate system, and $P_{N/S}$ is the vector directed from the solar center to the north/south pole (refer to Figure~\ref{fig:alphadiagram}). For a spacecraft orbiting at L4 (or L5) with $\alpha = 0^\circ$, $\mathbf{R}_{LOS}^{t}$ of its onboard imaging instrument can reach the highest solar latitude (both in the north and south) that is the orbital inclination plus the tilt of $7.25^\circ$ between the solar equator and the ecliptic plane. $\alpha$ is therefore set as $0^\circ$ in all our visibility analysis. Figure~\ref{fig:alpha_vs_lat} illustrates the time variation of the heliographic latitude for a given surface grid cell of which $\Theta^{t}_{ij}$ is $0^\circ$ when viewed from spacecraft orbiting at L4 (or L5) with the same orbital inclination of $14.5^\circ$ but deployed into the inclined orbit with a different set of $\alpha$ values.

\section{Results of Solar Visibility Analysis~\label{sec:results}}
%This section provides various visibility analyses of the solar surface considering both single and multiple spacecraft positions at L1, L4 and L5.
%Initially, the focus is on the on-disk observation from a single spacecraft situated along the Sun-Earth line and at the Sun-Earth triangular Lagrange point. We vary $\Theta_\texttt{max}$ to assess its impact on visibility for the spacecraft located along the Sun-Earth line. The spacecraft located in the triangular Lagrange point undergoes identical analysis but with varying inclinations to understand the advantages of inclined triangular Lagrange point's high-latitude observation.
%Subsequently, we extend our analysis to multiple spacecraft at L1, L4 and L5 to understand the synergistic effects while adjusting each spacecraft's $\Theta_\texttt{max}$. This step allows us to explore the combined visibility contributions from different vantage points.
%We conduct a limb observation simulation to validate the advantages of deploying multiple spacecraft at the triangular Lagrange points. Assuming a solar flare apex altitude of 10,000 km, we calculate the probability of observing solar flares (above M1.0) per year based on the coverage percentage of stereoscopic imaging, which considers regions covered by both on-disk and limb observations from multiple locations.
%Finally, we perform a visibility analysis of sunspots traversing the solar surface following the differential rotation. We consider different latitudes and vary $\Theta_\texttt{max}$ to evaluate the impact on sunspot visibility.

\subsection{Remote-sensing Observations from a Single Spacecraft~\label{subsec:results_single}}

\begin{figure}[htbp]
  \centering
  \includegraphics[width=.99\textwidth]{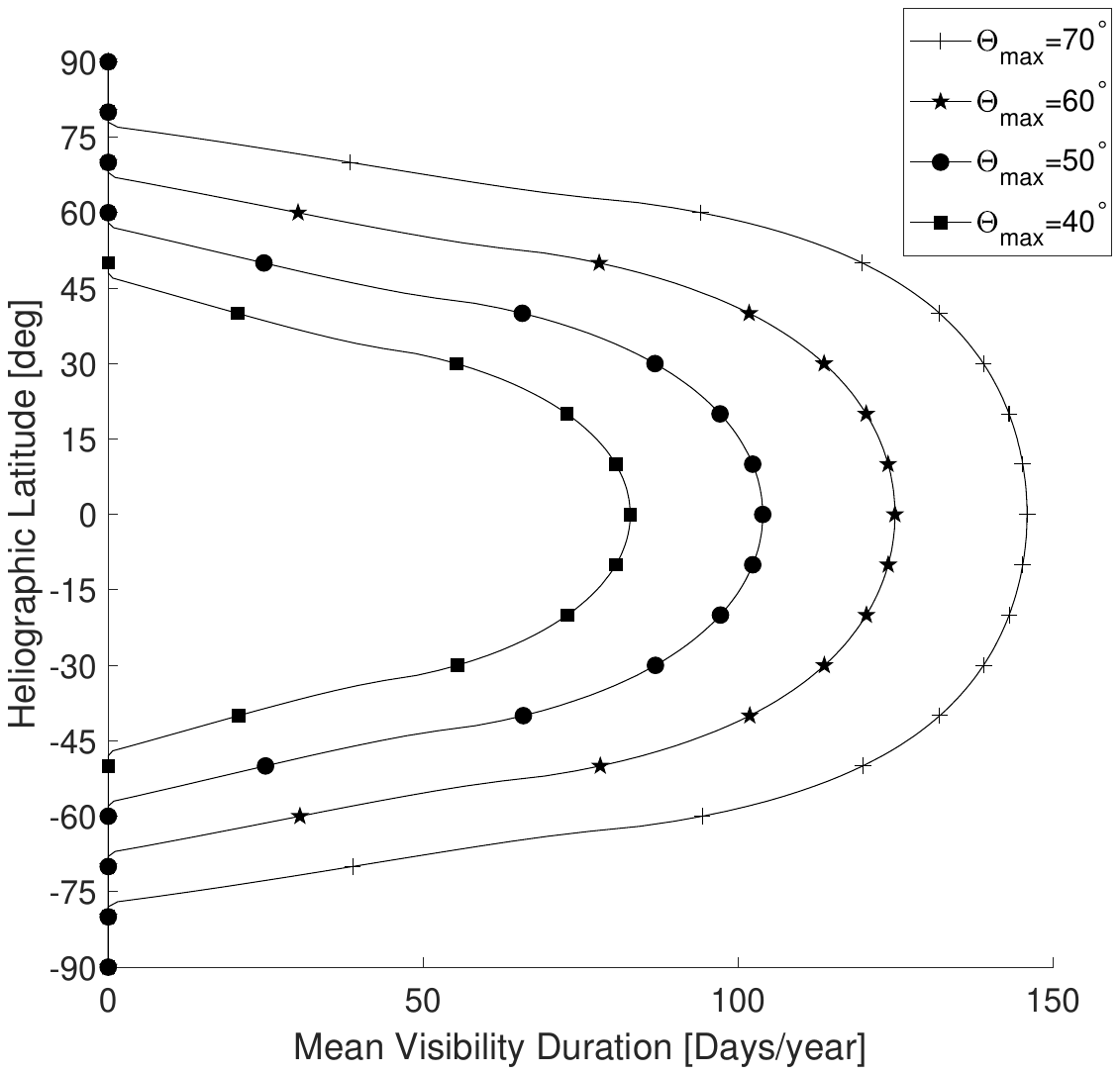}
  \caption{The mean visibility duration of the Sun's surface as viewed from spacecraft along the Sun-Earth line as a function of heliographic latitude. Different values of $\Theta_\texttt{max}$ are set to assess its impact on visibility.}
  \label{fig:L1_visibility}
  \centering
\end{figure}

%There have been few space missions off the Sun-Earth line to carry out remote-sensing observations of the Sun: e.g., the two nearly identical STEREO spacecraft (STEREO-A and STEREO-B) in Earth-like orbits around the Sun (gradually drifting away from Earth). Most of solar observations have been made on and near the Sun-Earth line, including ground-based telescopes and solar-observing satellites placed at LEO, GEO and L1. 
The mean visibility duration ($\bar{\Psi}_{i}$, in the units of days per year) of the solar surface as viewed from a single spacecraft is examined as a function of heliographic latitude. Figure~\ref{fig:L1_visibility} shows $\bar{\Psi}_{i}$ for spacecraft placed along the Sun-Earth line or ground telescopes on Earth's surface. With the orbital orientation $\alpha=0$, the visibility analysis is performed with different values of $\Theta_\texttt{max}=40^\circ$, $50^\circ$, $60^\circ$ and $70^\circ$, respectively, to assess its impact on the solar surface visibility. As expected, the larger $\Theta_\texttt{max}$ provides the longer visibility duration. 

\begin{figure}[htbp]
\centering
  \begin{tabular}{@{}cccc@{}}
    \begin{subfigure}[b]{0.5\linewidth}
        \centering
        \includegraphics[width=.99\textwidth]{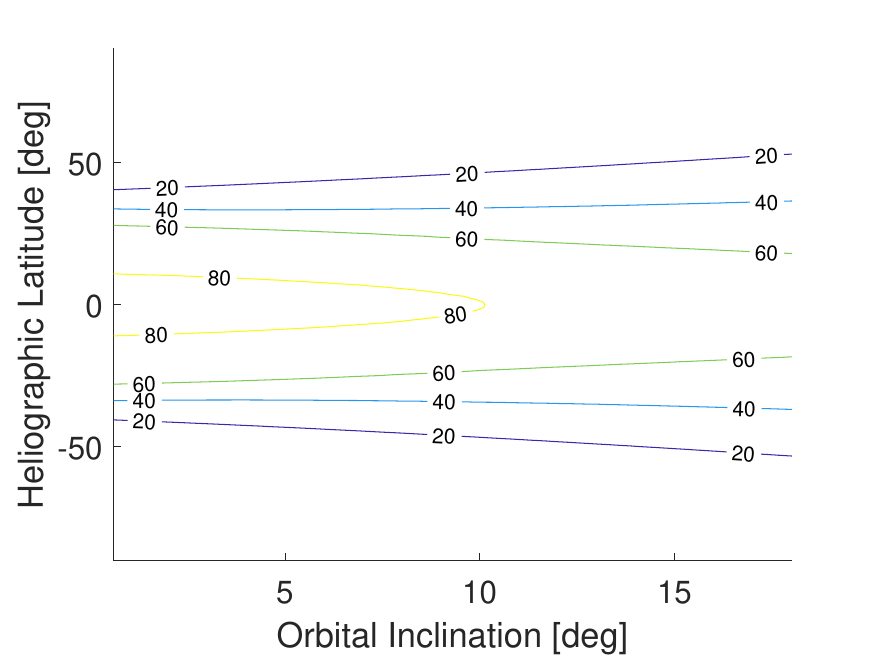}
        \caption{$\Theta_\texttt{max}=40$}\label{fig:a}
    \end{subfigure}%  
    \begin{subfigure}[b]{0.5\linewidth}
        \centering
        \includegraphics[width=.99\textwidth]{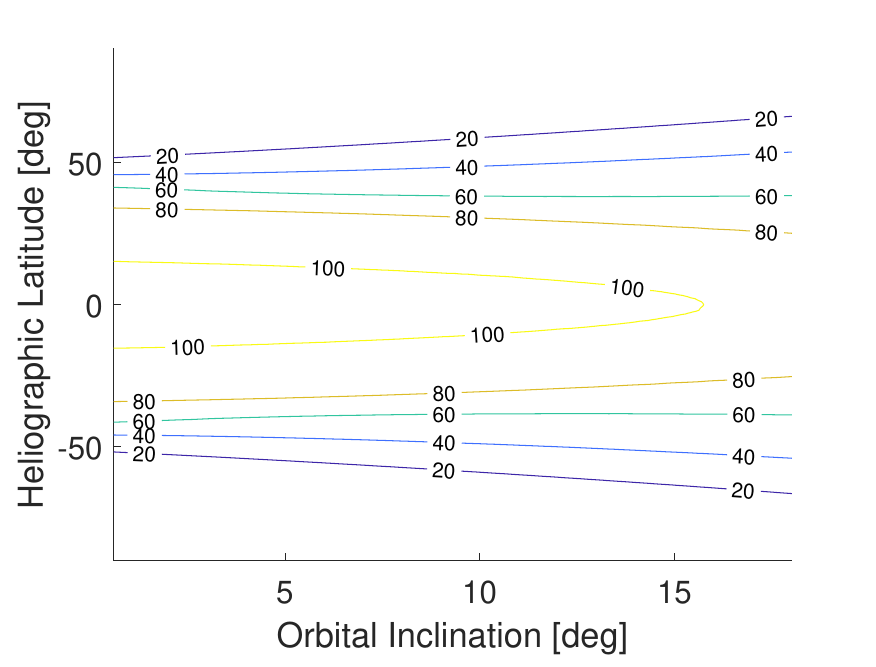}
        \caption{$\Theta_\texttt{max}=50$}\label{fig:b}
    \end{subfigure}\\
    \begin{subfigure}[b]{0.5\linewidth}
        \centering
        \includegraphics[width=.99\textwidth]{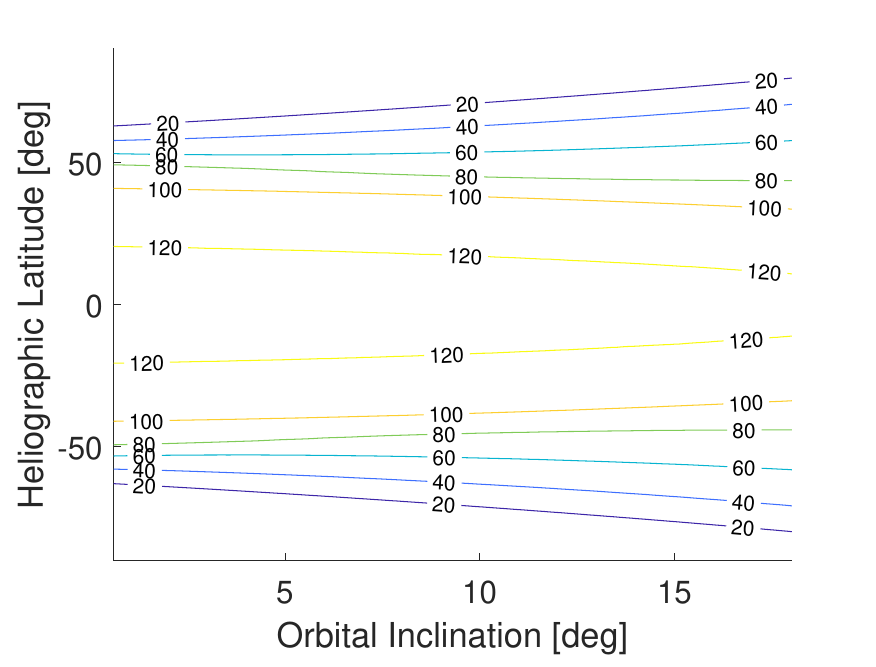}
        \caption{$\Theta_\texttt{max}=60$}\label{fig:c}
    \end{subfigure}%  
    \begin{subfigure}[b]{0.5\linewidth}
        \centering
        \includegraphics[width=.99\textwidth]{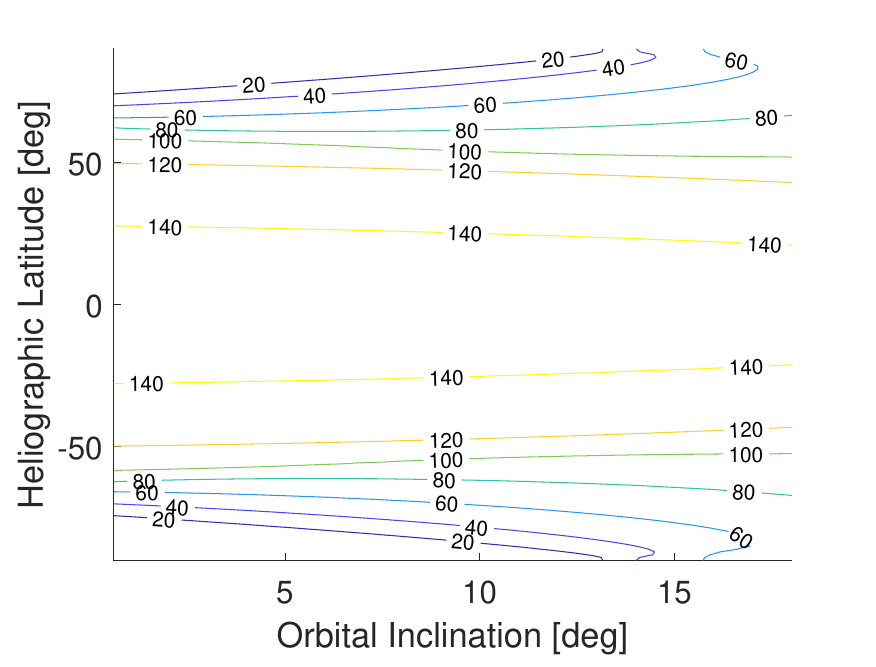}
        \caption{$\Theta_\texttt{max}=70$}\label{fig:d}
    \end{subfigure}%  
  \end{tabular}
  \caption{As viewed from spacecraft orbiting around L4 (or L5) with different inclinations, the mean visibility duration of the solar surface (marked as numbers on contour lines) is presented for a given heliographic latitude. $\Theta_\texttt{max}$ is set to $40^\circ$ (panel a), $50^\circ$ (panel b), $60^\circ$ (panel c) and $70^\circ$ (panel d), respectively.}
  \label{fig:thetavsinc}
\end{figure}

Compared to spacecraft at L1, vertical periodic orbits at L4 and L5 can have a large inclination, as demonstrated in Figure~\ref{fig:L4PO}. For a given   $\Theta_\texttt{max}$, Figure~\ref{fig:thetavsinc} shows $\bar{\Psi}_{i}$ for a single spacecraft placed at L4 (or L5) with different orbital inclinations. A spacecraft orbiting around L4 (or L5) with a larger inclination is able to observe higher latitudes for an extended period. This however comes at the expense of a shorter observation duration for latitudes closer to the Sun's equator. This phenomenon is a natural outcome of the spacecraft's oscillatory motion in the solar surface's latitudinal direction. 
The visibility duration characteristics remain consistent as $\Theta_\texttt{max}$ increases until $\Theta_\texttt{max}=70^\circ$ where the observation trend deviates. As depicted in Figure~\ref{fig:d}, a spacecraft at the triangular Lagrange point with an $18^\circ$ inclination can observe latitudes of $90^\circ$ for a longer duration than latitudes between $70^\circ $ and $ 80^\circ$. This phenomenon arises from the spacecraft's capability to observe beyond the pole.
Through this analysis, critical mission requirements and constraints can be determined. For instance: 1) given $\Theta_\texttt{max}$ (from an optical camera) and the desired solar surface latitude to observe, what should be the inclination of the L4 periodic orbit to achieve observation for "X" number of days per year? and 2) given the inclination of the L4 periodic orbit, what is the expected data quality at the desired latitude of the solar surface?

\subsection{Synergy of Multi-Point On-Disk Observation\label{subsec:results_multipoint}}

Multi-point observation has become a pivotal objective for the heliospheric science division of space agencies. Various white papers (\cite{posner2021multi}; \cite{gopalswamy2023multiview}; \cite{posner2023focused}; \cite{han2023unveiling}; \cite{cho2023opening}) have been authored to propose the design of a multi-view observatory spacecraft system situated near the triangular Lagrange point and the first Lagrange point. This configuration aims to enable access to a larger longitude range of the solar surface for an extended duration and from multiple angles. As indicated by \cite{gopalswamy2023multiview}, observing the solar surface from multiple angles offers insights to address numerous fundamental questions in heliophysics research. Some of these unanswered questions include:
\begin{enumerate}
  \item How do active regions evolve before and after emerging to the solar surface? 
  \item How do coronal mass ejections (CME) flux ropes form, accelerate, drive shocks, and evolve from near the Sun into the heliosphere, including particle acceleration? 
  \item How do corotating interaction regions (CIR) magnetic fields evolve in the inner heliosphere and accelerate particles? 
  \item How does shock geometry and magnitude evolve and how does this relate to solar energetic particles (SEPs) and radio bursts?
  \item Can multi-point active region observations aid in forecasting their activity?
\end{enumerate}

Placing spacecraft at the triangular Lagrange points provides access to the solar region behind the solar limb of the spacecraft situated in Sun-Earth L1. This strategic positioning becomes especially valuable in studying phenomena such as CME-expelled particles that accelerate through space following the Sun's magnetic field lines. If a spacecraft was located in L4, it could effectively measure and detect potential dangers in the space environment for missions, such as manned Lunar and Mars missions. As outlined in \cite{posner2021multi}, the additional flare information from Sun-Earth L4 would enable gathering sufficient data to comprehensively cover all Solar Energetic Particles (SEPs) that could significantly impact missions in the Earth-Moon system.

\begin{figure}[htbp]
\centering
  \begin{tabular}{@{}cccc@{}}
    \begin{subfigure}[b]{0.5\linewidth}
        \centering
        \includegraphics[width=.99\textwidth]{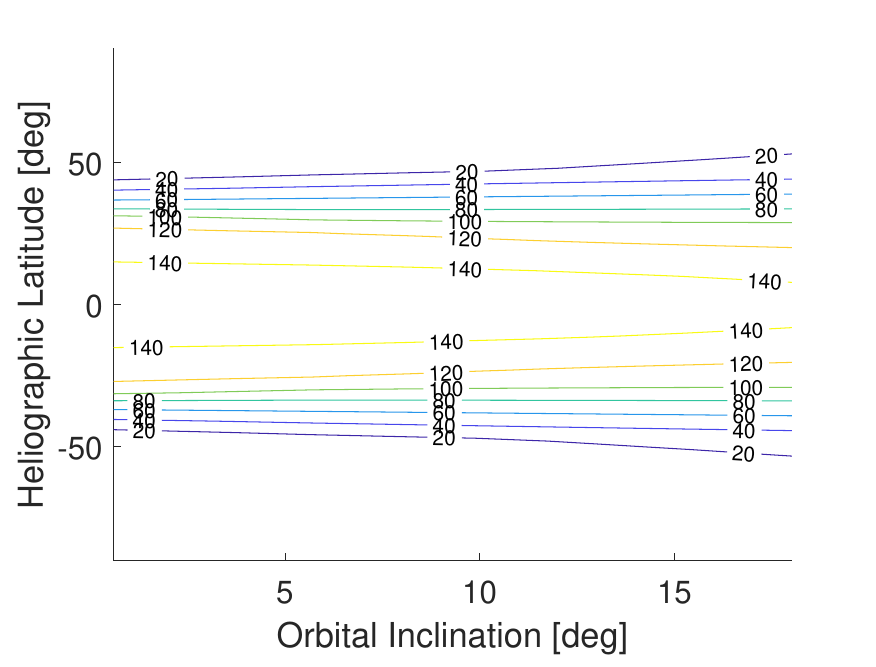}
        \caption{$\Theta_\texttt{max}=40$}
    \end{subfigure}%  
    \begin{subfigure}[b]{0.5\linewidth}
        \centering
        \includegraphics[width=.99\textwidth]{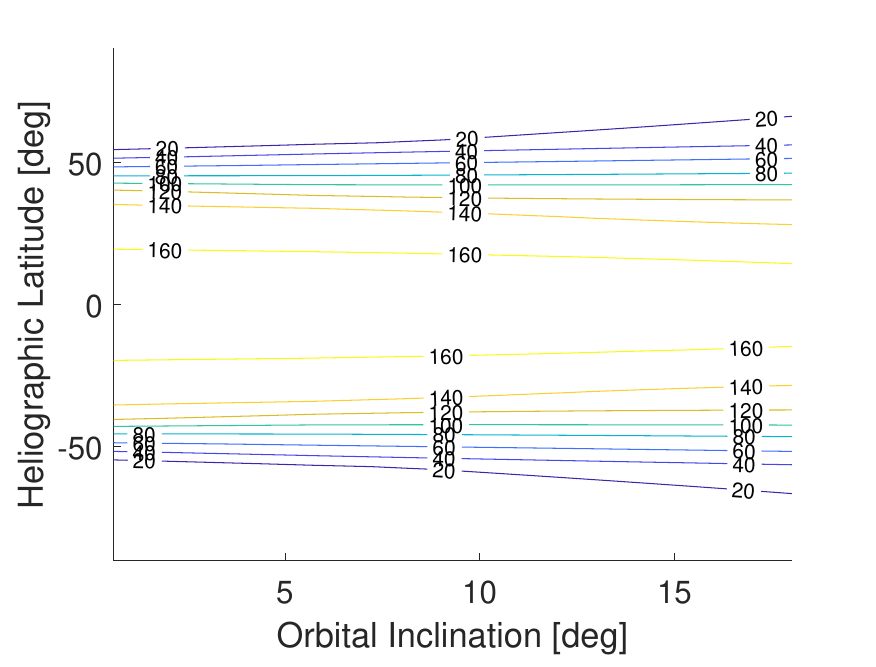}
        \caption{$\Theta_\texttt{max}=50$}
    \end{subfigure}\\
    \begin{subfigure}[b]{0.5\linewidth}
        \centering
        \includegraphics[width=.99\textwidth]{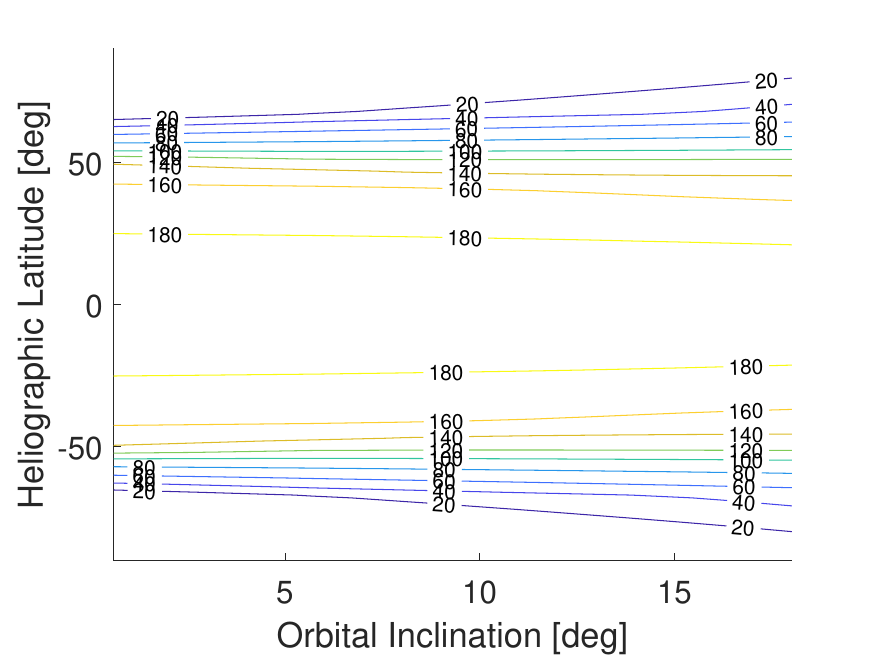}
        \caption{$\Theta_\texttt{max}=60$}
    \end{subfigure}%  
    \begin{subfigure}[b]{0.5\linewidth}
        \centering
        \includegraphics[width=.99\textwidth]{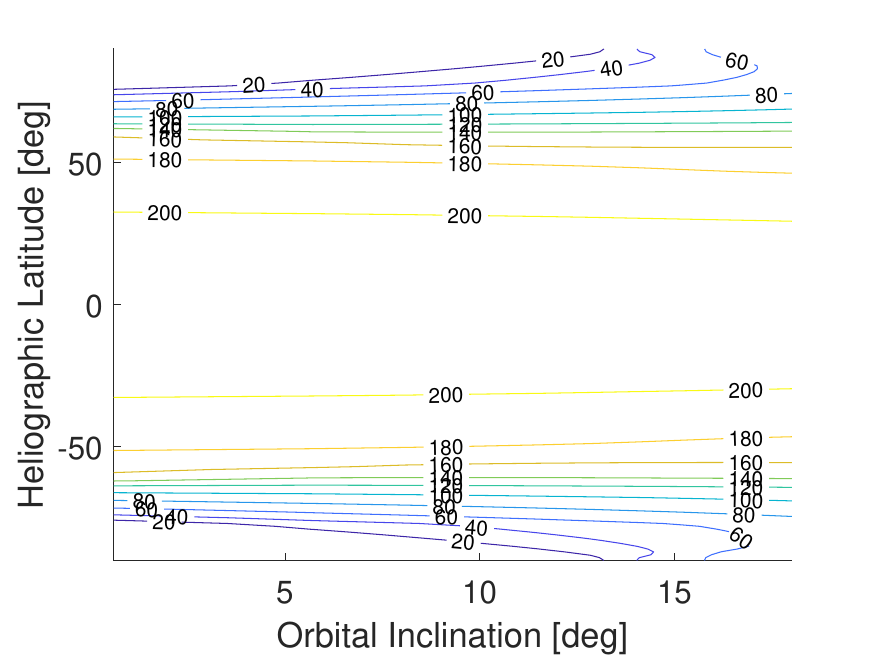}
        \caption{$\Theta_\texttt{max}=70$}
    \end{subfigure}%  
  \end{tabular}
  \caption{Visibility analysis of solar surface for different latitudes ($\phi$) with a varying inclination ($i$) of triangular Lagrange points and $\Theta_\texttt{max}$ using one spacecraft at L1, L4 and L5 (total three spacecraft). Number marks on the contour lines show the observation duration measured in days/year.}
  \label{fig:Multipointsinglesat}
\end{figure}

The improvement in visibility duration resulting from placing additional spacecrafts at L4 and L5 with $\Theta_\texttt{max}=60^\circ$ are presented in Table \ref{tab:improvements_ondisk_zeroinc} and \ref{tab:improvements_ondisk_145inc}, for L4 and L5 spacecraft in planar-Lyapunov orbits and 14.5 inclined triangular periodic orbits, respectively (retrieved from Figure \ref{fig:Multipointsinglesat}). The most significant differences occur between triangular Lagrange point periodic orbits' inclinations for high latitude observation. Spacecraft placed in L4 periodic orbit without inclination doubles the visibility duration for high latitudes. However, spacecraft placed in L4 periodic orbit with a 14.5-degree inclination about the ecliptic with $\alpha=0$ increases the visibility duration by 131\%. When considering the L5 spacecraft, the improvements become even more substantial (268\%).

\begin{table}[htbp]
    \centering
    \caption{Solar surface visibility improvements for L1+L4 and L1+L4+L5 planar Lyapunov orbit with $\Theta_\texttt{max} = 60^\circ$ compared to a single spacecraft at L1. Integers cell defines number of averaged visible days per year.}
    \begin{tabular}{ccccc}
        \hline
        Spacecraft Locations & $\phi=0^\circ$ & $\phi=20^\circ$ & $\phi=40^\circ$ & $\phi=60^\circ$\\
        \hline
        L1 (days) & 124 & 120 & 110 & 32\\
        L1+L4 (days) & 187 & 182 & 164 & 60\\
        Imp. (\%) & 51\% & 52\% & 64\% & 87.5\%\\
        L1+L4+L5 (days) & 250 & 244 & 228 & 90\\
        Imp.(\%) & 101\% & 103\% & 128\% & 181\%\\
        \hline
    \end{tabular}
    \label{tab:improvements_ondisk_zeroinc}
\end{table}

\begin{table}[htbp]
    \centering
    \caption{Solar surface visibility improvements for L1+L4 and L1+L4+L5 (planar Lyapunov orbit for L1 and 14.5 inclination for L4 and L5 periodic orbits) with $\Theta_\texttt{max} = 60^\circ$ compared to a single spacecraft at L1. Integers cell defines number of averaged visible days per year.}
    \begin{tabular}{ccccc}
        \hline
        Spacecraft Locations & $\phi=0$ & $\phi=20$ & $\phi=40$ & $\phi=60$\\
        \hline
        L1 (days)& 124 & 120 & 110 & 32\\
        L1+L4 (days) & 186 & 181 & 153 & 74\\
        Imp. (\%)& 50\% & 51\% & 53\% & 131\%\\
        L1+L4+L5 (days)& 247 & 242 & 210 & 118\\
        Imp.(\%)& 99\% & 101\% & 110\% & 268\%\\
        \hline
    \end{tabular}

    \label{tab:improvements_ondisk_145inc}
\end{table}

\begin{figure}[htbp]
\centering
  \begin{tabular}{@{}cccc@{}}
    \begin{subfigure}[b]{0.5\linewidth}
        \centering
        \includegraphics[width=.99\textwidth]{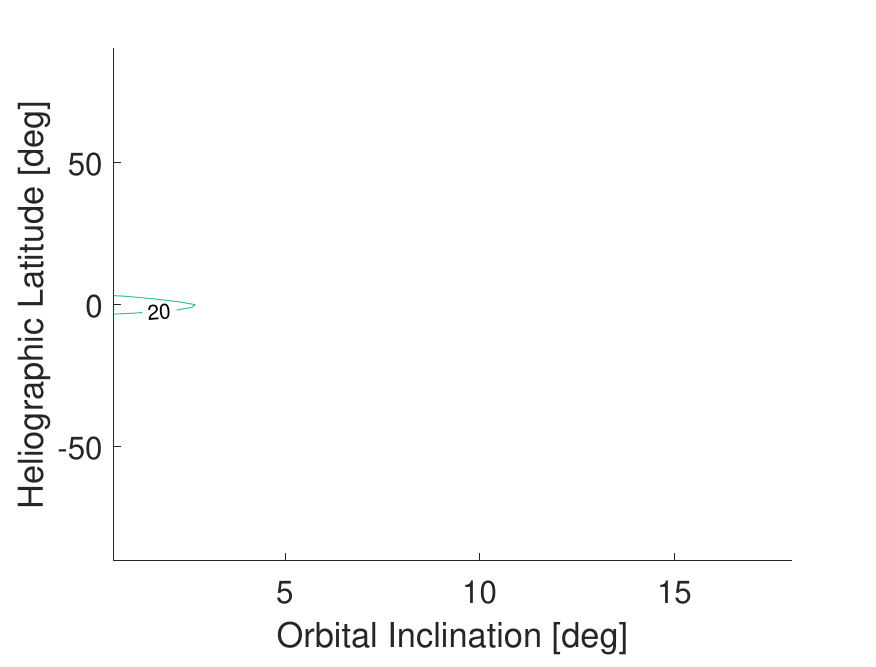}
        \caption{$\Theta_\texttt{max}=40$}
    \end{subfigure}%  
    \begin{subfigure}[b]{0.5\linewidth}
        \centering
        \includegraphics[width=.99\textwidth]{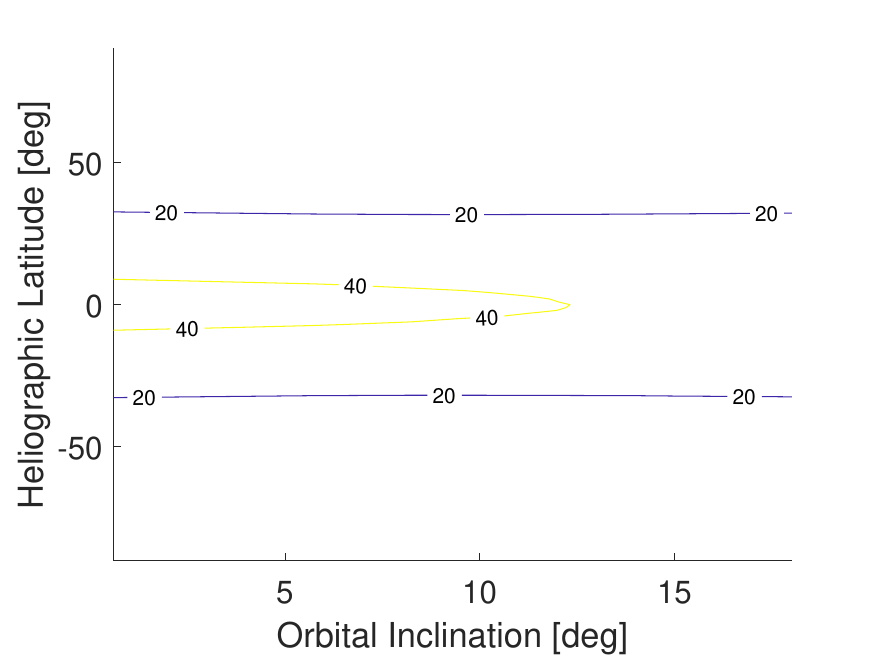}
        \caption{$\Theta_\texttt{max}=50$}
    \end{subfigure}\\
    \begin{subfigure}[b]{0.5\linewidth}
        \centering
        \includegraphics[width=.99\textwidth]{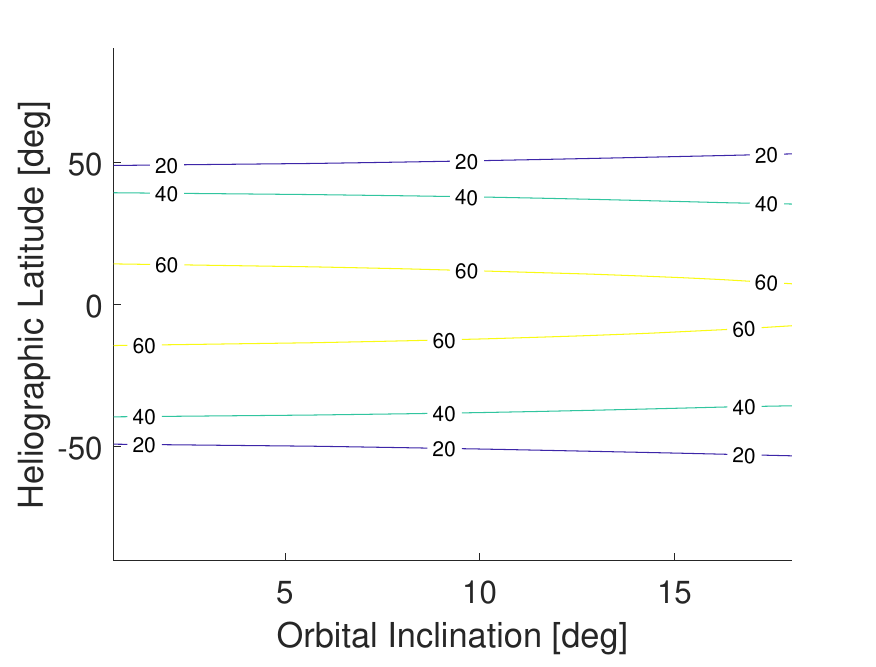}
        \caption{$\Theta_\texttt{max}=60$}
    \end{subfigure}%  
    \begin{subfigure}[b]{0.5\linewidth}
        \centering
        \includegraphics[width=.99\textwidth]{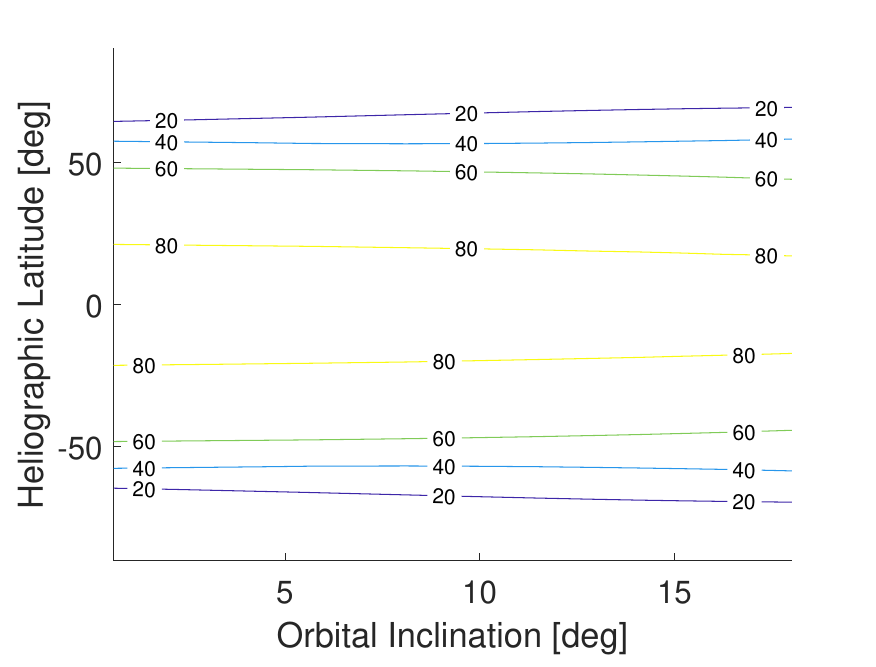}
        \caption{$\Theta_\texttt{max}=70$}
    \end{subfigure}%  
  \end{tabular}
  \caption{Visibility analysis of the solar surface with a multi-point view for different latitudes ($\phi$) with a varying inclination{$i$} of triangular Lagrange points and $\Theta_\texttt{max}$ using one spacecraft at L1, L4 and L5. Number marks on the contour lines show the observation duration measured in days/year.}
  \label{fig:multipointmultisat}
\end{figure}

A similar analysis is performed for multi-point, on-disk observation, which measures the visible duration of a specific latitude observed by two spacecraft. Figure~\ref{fig:multipointmultisat} illustrates the impact of varying $\Theta_\texttt{max}$ and the inclination of the periodic orbit on the observation duration of multi-spacecraft placed in L1, L4 and L5. Unlike Figure~\ref{fig:Multipointsinglesat}, the multi-spacecraft view constraint (visible by two spacecraft at the same time) exhibits significant differences with respect to the value of $\Theta_\texttt{max}$. This phenomenon arises due to the inherent nature of the viewing angle differences between L1, L4 and L5, which are separated by $60^\circ$.
Due to such separation angle, the overlapping of solar surface on-disk observation begins at $\Theta_\texttt{max} = 30$, and the overlapping increases as $\Theta_\texttt{max}$ increases.

\subsection{Synergy of Multi-Point Solar Limb Observation\label{subsec:results_limb}}
Observing the solar limb is critical in reconstructing the solar magnetic field. While on-disk observation provides constraints on the field of the photosphere, solar limb observation data can add constraints on overlying coronal magnetic field structure. This may be crucial for solar magnetic field reconstruction, particularly leading up to solar flares. Unfortunately, up to recently, there has not been an instance where a solar flare was observed simultaneously from both the solar limb and on-disk, as reported by \cite{chen2020measurement} and \cite{fleishman2020decay}. \cite{chen2020measurement} managed to observe on-disk activity four days prior, while \cite{fleishman2020decay} observed it two days before the solar flare occurred near the limb. A multi-spacecraft mission strategically positioned at L1, L4 and L5 holds the potential to significantly increase the frequency of observing the solar limb from multiple angles, and from the limb. This advantage stems from the unique geometry of the triangular Lagrange points with respect to the Sun-Earth line. 

\begin{figure}[htbp]
  \centering
  \includegraphics[page=10,scale=0.8]{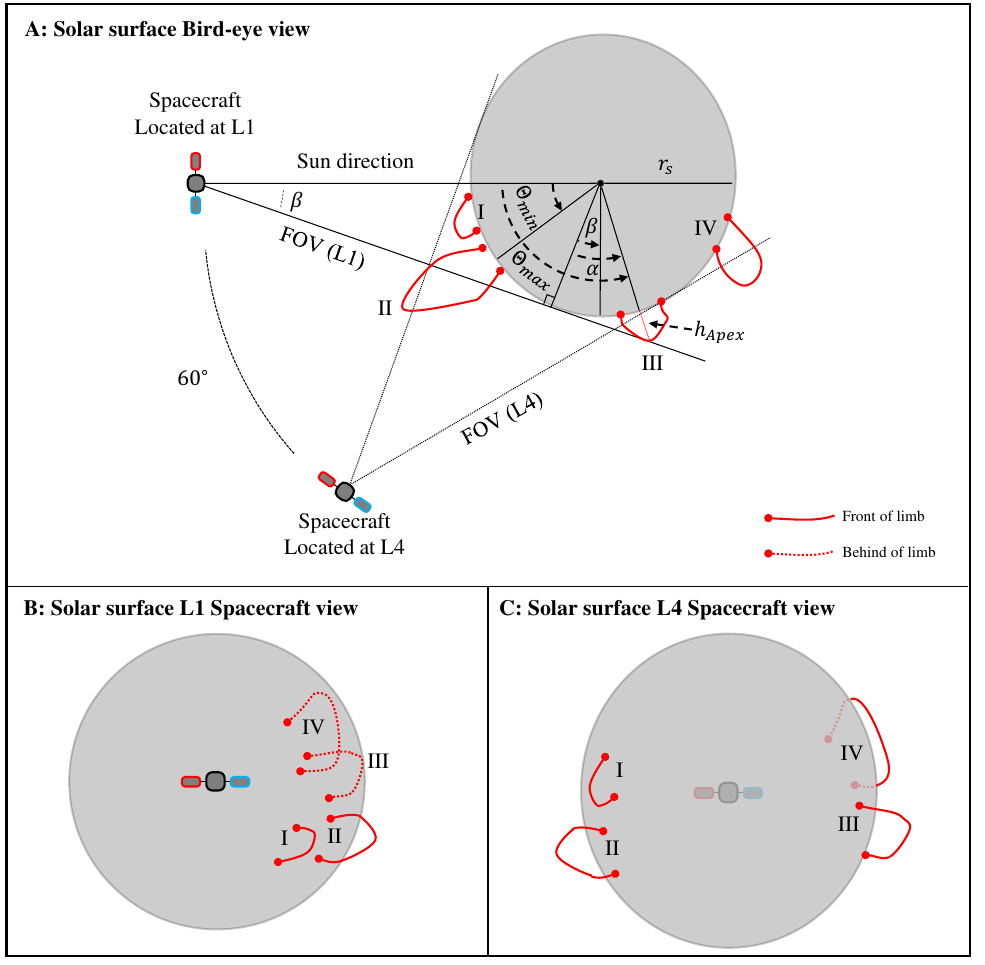}
  \caption{Top and spacecraft's view of solar flares erupting from L1 and L5 of the solar surface.}
  \label{fig:solarFlareApex}
\end{figure}

The visibility condition for solar flares is determined based on the flare loop apex. Figure \ref{fig:solarFlareApex} shows the geometry utilized to determine the solar flare visibility. The figure shows four possible occurrences of the solar flares seen from both bird-eye and spacecraft viewpoints from L1 and L4. From the perspective of spacecraft located at L1, flare $I$ represents solar flare occurring within the on-disk observation of the Sun from the spacecraft, which processes large projection effects from the solar surface behind the solar flares. We assume the flare $I$ cannot be used as scientific data. Flare $II$ represents solar flare with the loop-top source outside the disk but with the footprints inside the on-disk observation. Flare $III$ and $IV$ represent an invisible solar flare (both footprints and loop-top) as both features are on the other side of the limb. We consider the solar flares with the loop-top visible outside of the solar limb as visible. Hence, solar flare $II$ is considered visible, whereas solar flares $I$, $III$ and $IV$ are considered invisible from L1.

\begin{figure}[htbp]
  \centering
  \includegraphics[page=9,scale=0.175]{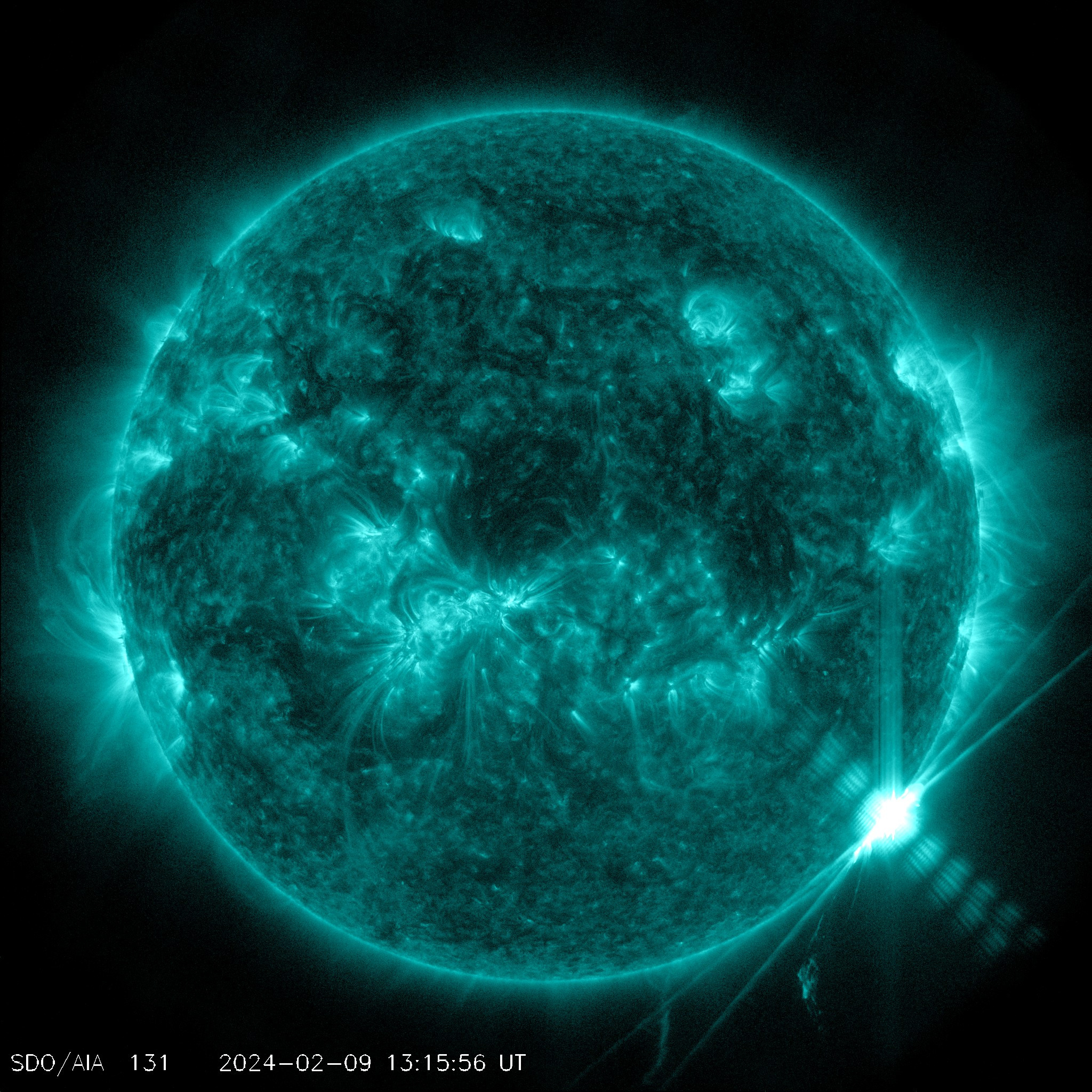}
  \caption{An enormous flare on Feb. 9, 2024, from behind the west limb (W110). The visible flare was magnitude X6.2 with occulted foot points (Case III in Figure~\ref{fig:solarFlareApex}, image taken from Solar Dynamics Observatory)}
  \label{fig:solarFlareApex_february}
\end{figure}

However, when the identical solar flares are viewed simultaneously from spacecraft located in L4, different features of the solar flares are observable, as shown in Figure~\ref{fig:solarFlareApex}. Except for solar flare $I$, which is still invisible due to the low solar flare loop-top apex, flare $II$, $III$, and $IV$ are observable. Solar flare $IV$, which was invisible from the spacecraft at L1, is visible from L4 because of the separation angle of L1 and L4. Flare $IV$ represents behind-the-limb solar flare loop-top observation with occulted footprints as the footprints are on the other side of the solar limb. An example of Flare $IV$ can be seen in figure~\ref{fig:solarFlareApex_february}. Additionally, the solar flares $II$ and $III$ are observed from different angles, allowing the stereoscopic observation and advancing the 3D reconstruction of the solar flares.

Based on the geometry shown in Figure~\ref{fig:solarFlareApex}, the minimum and maximum angle of visibility for solar flares with averaged solar flare loop apex of $10,000km$ (\cite{reale1997determination}, \cite{reale2014coronal}) are 80.08 and 99.39 for $\Theta_\texttt{min}$ and $\Theta_\texttt{max}$, respectively. $\Theta_\texttt{min}$ and $\Theta_\texttt{max}$ is calculated as follows
\begin{equation}
    \Theta_\texttt{min} = \frac{\pi}{2} - \alpha - \beta
\end{equation}
\begin{equation}
    \Theta_\texttt{min} = \frac{\pi}{2} + \alpha - \beta
\end{equation}
where $\alpha$ and $\beta$ is calculated as follows
\begin{equation}
    \alpha = \cos^{-1}\left(  \frac{r_s}{r_s+h_{apex}} \right)
\end{equation}
\begin{equation}
    \beta = \sin^{-1}\left(  \frac{r_s}{AU} \right)
\end{equation}
where $h_{apex}$ defines the altitude of the solar flare apex, $r_s$ defines the radius of the Sun, and lastly, the AU defines the astronomical unit, tabulated in table \ref{tab:nond}. The solar surface coverage with varying colors is plotted in section \ref{sec:appendix}. 

According to \cite{schrijver2012estimating}, the average number of solar flare events with energies larger than M1.0 is 140 per year, based on the events recorded from 1996 to 2007 (one solar cycle). Additionally, it is recorded that most of the solar flare events occur within the $N40^\circ$ and $S40^\circ$. From the simulations plotted in the Appendix section, the solar surface coverage for stereoscopic coverage is derived as $11.8\%$ and $24\%$ for dual-spacecraft (L1, L4(or L5)) and triple-spacecraft (L1, L4 and L5) mission, respectively, throughout the year. Therefore, we can conclude that under the two assumptions: 1) all solar flares have $h_{apex}$ higher than 10,000km, and 2) all solar flares occur within $N40^\circ$ and $S40^\circ$, placing additional spacecraft at L4 and L5 will provide multiple-angle on-disk observation and limb observation of $\sim34$ flares per year (\cite{reale1997determination}, \cite{reale2014coronal}).

\subsection{Multi-Point Continuous Sunspot Observation\label{subsec:results_sunspot}}

\begin{figure}[htb]
  \centering
  \includegraphics[scale=0.7]{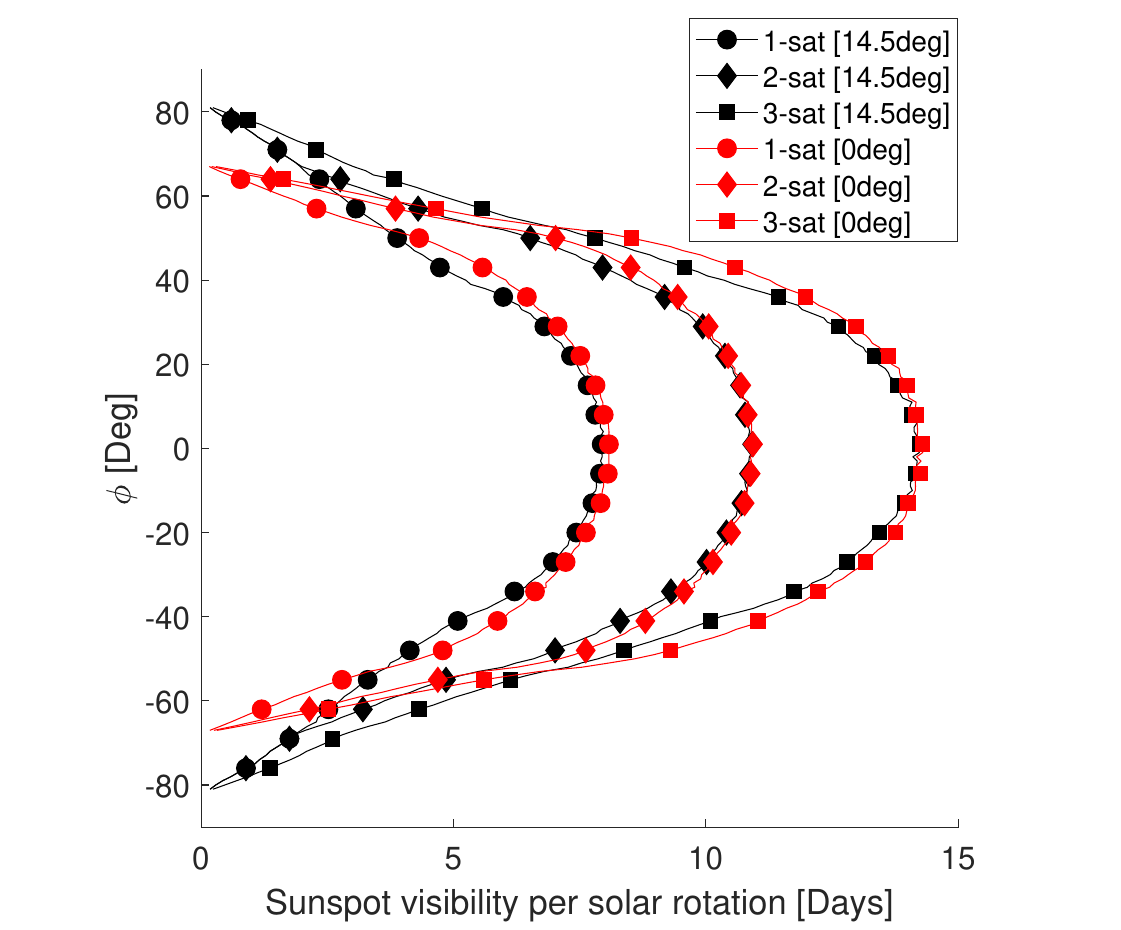}
  \caption{Sunspot observation per sidereal differential rotation for different latitude with  $\Theta_\texttt{max} = 60^\circ$}
  \label{fig:sunspot}
\end{figure}

Observing sunspots is essential for advancing solar magnetic field research. However, observing the long-term evolution of sunspots, from emergence and development to decay, has been challenging due to the lack of on-disk coverage of the solar surface. This section aims to elucidate the advancements in the capability of observing sunspots at different latitudes with a predefined $\Theta_\texttt{max}$ of $60^\circ$ for spacecraft pairs of L1+L4 and L1+L4+L5, both with planar and vertical periodic orbits.
Figure~\ref{fig:sunspot} depicts the sunspot visibility in days per solar rotation, which rotates with a differential rotation speed as given by Eq. \ref{differentialrotationEQ}, with $\Theta_\texttt{max}$ of $60^\circ$.
The improvements from placing multiple spacecraft at the triangular Lagrange points are tabulated in Tables \ref{tab:improvements_sunspot_0inc} and \ref{tab:improvements_sunspot_145inc} for $0^\circ$ and $14.5^\circ$ inclined triangular Lagrange points, respectively. As expected, sunspot visibility per solar rotation near the equator of the sun experiences a $1.5$ times improvement for each spacecraft added at L4 and L5, inherent to the geometry of the triangular Lagrange points with $\Theta_\texttt{max}=60^\circ$. Significant improvements occur near high-latitude sunspots when inclined periodic orbits are considered. Compared to placing L4 and L5 spacecraft in a planar orbit, the vertical periodic orbit option improves sunspot visibility per solar rotation at $\phi=60^\circ$ by 200\%. Interestingly, inclined triangular Lagrange points exhibit smaller improvement in sunspot visibility per solar rotation for lower solar surface latitudes ($\phi=0, 20, 40$). This is due to the nature of the inclined triangular Lagrange point's latitudinal oscillation motion with respect to the ecliptic plane.

\begin{table}[htb]
    \centering
    \caption{Sunspot visibility improvements for L1+L4 and L1+L4+L5 planar Lyapunov orbit with $\Theta_\texttt{max} = 60^\circ$ compared to a single spacecraft at L1 in sunspot visibility days per solar rotation.}
    \begin{tabular}{ccccc}
        \hline
        Spacecraft Locations & $\phi=0$ & $\phi=20$ & $\phi=40$ & $\phi=60$\\
        \hline
        L1 (days)& 8.0 & 7.52 & 5.92 & 1.6\\
        L1+L4 (days)& 11.2 & 10.56 & 8.96 & 2.72\\
        Imp. (\%) & 40\% & 40.5\% & 51\% & 70\%\\
        L1+L4+L5 (days)& 14.4 & 13.76 & 11.20 & 3.36\\
        Imp.(\%) & 80\% & 82\% & 89\% & 110\%\\
        \hline
    \end{tabular}
    
    \label{tab:improvements_sunspot_0inc}
\end{table}

\begin{table}[htb!]
    \centering
    \caption{Sunspot visibility improvements for L1+L4 and L1+L4+L5 (planar Lyapunov orbit for L1 and 14.5 inclination for L4 and L5 periodic orbits) with $\Theta_\texttt{max} = 60^\circ$ compared to a single spacecraft at L1 in sunspot visibility days per solar rotation.}
    \begin{tabular}{ccccc}
        \hline
        Spacecraft Locations & $\phi=0$ & $\phi=20$ & $\phi=40$ & $\phi=60$\\
        \hline
        L1 (days)& 8.0 & 7.52 & 5.92 & 1.6\\
        L1+L4 (days)& 10.88 & 10.4 & 8.48 & 3.52\\
        Imp. (\%) & 36\% & 38\% & 43\% & 120\%\\
        L1+L4+L5 (days)& 14.34 & 13.44 & 10.24 & 4.8\\
        Imp.(\%) & 78\% & 79\% & 72\% & 200\%\\
        \hline
    \end{tabular}
    
    \label{tab:improvements_sunspot_145inc}
\end{table}

\newpage

\section{Summary and discussion ~\label{sec:summary}}

In this study, we conducted a comprehensive analysis of solar surface visibility for a multi-spacecraft mission, aiming to quantify the advantages and disadvantages of various orbit configurations in the Sun-Earth system's Lagrange points. Our approach involved using the L1 Lyapunov orbit and planar Lyapunov orbit for spacecraft located at L1, and both inclined and planar periodic orbits for the triangular Lagrange points. For a single spacecraft, our analysis showed that placing a spacecraft at the Sun-Earth L4 planar orbit increased on-disk observation by 51\% and 87.5\% for latitudes $\phi=0^\circ$ and $\phi=60^\circ$, respectively. Introducing a spacecraft at Sun-Earth L4 with a $14.5^\circ$ inclination about the ecliptic plane expanded the observation duration by 131\% compared to a single spacecraft at Sun-Earth L1. Furthermore, identical spacecraft placed at L5, mirroring the inclination of the L4 spacecraft, experienced a 268\% improvement in observing $\phi=60^\circ$, compared to a single spacecraft at Sun-Earth L1. The limb observation, lying within the range of W20 to W40 of the Carrington rotating frame due to the geometry of the triangular Lagrange points, aligns with on-disk observations from L1 and L4 for $\Theta_{\texttt{max}} > 30^\circ$. This alignment enhances the potential for magnetic field reconstruction by combining coronal constraints from limb observations with on-disk magnetograph observations. Regarding sunspot observation, the duration significantly increases with additional spacecraft at the triangular Lagrange points. However, improvements are more substantial in high-latitude observations, with smaller gains in lower altitudes due to the latitudinal oscillation of inclined triangular Lagrange points with respect to the ecliptic plane. The implications of these findings are vast for future heliospheric sciences and human missions to the Moon and Mars. A multi-spacecraft mission at L4 and L5 enhances our understanding of the Sun and may enable solar activity forecasting. This information is crucial for spacecraft safety in Earth orbit and future human missions beyond, providing unprecedented advantages in mitigating potential dangers posed by solar activity.

\begin{NiceTabular}{p{2cm} p{3cm} p{3cm} p{3cm}}[hvlines,colortbl-like]
\textbf{P=planar} \newline \textbf{I=Inclined}& \textbf{Solar surface feature coverage} & \textbf{High-latitude coverage} & \textbf{Limb observation} \\
        \hline
        \textbf{L1} & 
        \cellcolor{red!20} Limited around the Sun-Earth line.  &
        \cellcolor{red!20}Limited high-latitude coverage. &
        \cellcolor{red!20}Limb observation from L1.  \\
  
        \textbf{L1 + L4(P)} &
        \cellcolor{yellow!20}Elongated longitude coverage. Multi-point view starts from $\Theta_{\texttt{max}}= 30^\circ$. Sunspot observation duration increases.&
        \cellcolor{red!20} No increase in high-latitude coverage as L1 and L4 is on the ecliptic plane. &
        \cellcolor{yellow!20}Limb observation from L1 and L4. \\
 
        \textbf{L1 + L4(I)} &
        \cellcolor{yellow!20} Same as L1 + L4(P). Sunspot observation duration increases. High-latitude sunspots are visible with less projection effect. &
        \cellcolor{yellow!20}Modest increase in high-latitude coverage due to inclined L4 periodic orbit. &
        \cellcolor{yellow!20}Same as L1 + L4(P) \\

        \textbf{L1 + L4(P) + L5(P)} &
        \cellcolor{green!20}Larger longitude coverage. Significant increase in sunspot observation&
        \cellcolor{red!20}No increase in high-latitude coverage as L4 and L5 are on the ecliptic plane. &
        \cellcolor{green!20}Additional limb observation from L5. Multi-point view of L1/4 and L1/5 are covered by limb observation from L5 and L4, respectively. \\
 
        \textbf{L1 + L4(I) + L5(P)} &
        \cellcolor{green!20}Same as L1 + L4(P) + L5(P). High-latitude sunspots are visible with less projection effect. &
        \cellcolor{yellow!20}Modest increase in high-latitude coverage due to inclined L4 periodic orbit. &
        \cellcolor{green!20}Same as L1 + L4(P) + L5(P). \\

        \textbf{L1 + L4(I) + L5(I)} &
        \cellcolor{green!20}Same as L1 + L4(P) + L5(P). High-latitude sunspots are more frequently visible with less projection effect.&
        \cellcolor{green!20}Significant increase in high-latitude coverage duration. &
        \cellcolor{green!20}Same as L1 + L4(P) + L5(P). 
\end{NiceTabular}

\section{Appendix ~\label{sec:appendix}}
\begin{figure*}[htb]
\centering
  \begin{tabular}{@{}cc@{}}
    \begin{subfigure}[b]{.7\textwidth}
        \centering
        \includegraphics[page =1, width=.99\textwidth]{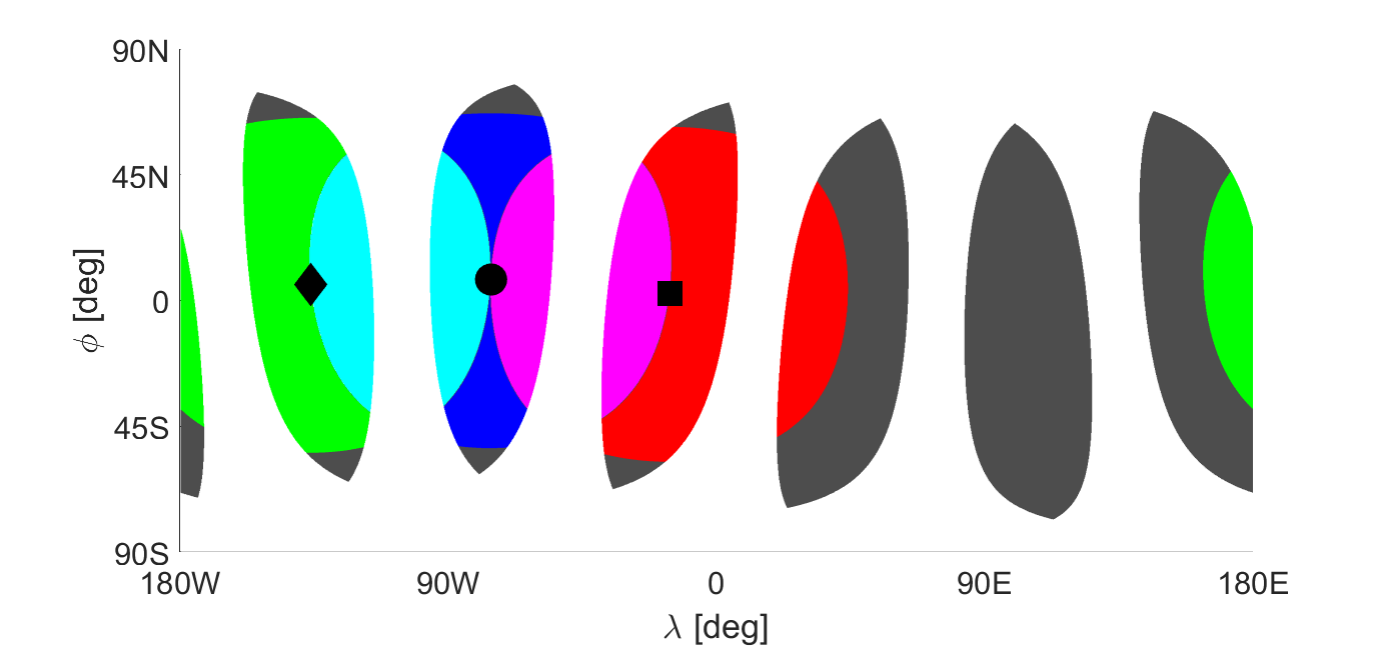}
        \caption{January}
    \end{subfigure}\\%  
    \begin{subfigure}[b]{.7\textwidth}
        \centering
        \includegraphics[page =2, width=.99\textwidth]{Figures/coverage.pdf}
        \caption{April}
    \end{subfigure}\\
    \begin{subfigure}[b]{.7\textwidth}
        \centering
        \includegraphics[page =3, width=.99\textwidth]{Figures/coverage.pdf}
        \caption{July}
    \end{subfigure}\\
   \begin{subfigure}[b]{.7\textwidth}
        \centering
        \includegraphics[page =4, width=.99\textwidth]{Figures/coverage.pdf}
        \caption{November}
    \end{subfigure}
  \end{tabular}
  \caption{Solar surface coverage with spacecraft located in planar L1, L4 and L5 with $\Theta_\texttt{max}=60^\circ$.}
  \label{fig:coveragemap_0}
\end{figure*}

\begin{figure*}[htb]
\centering
  \begin{tabular}{@{}cc@{}}
    \begin{subfigure}[b]{.7\textwidth}
        \centering
        \includegraphics[page=5,width=.99\textwidth]{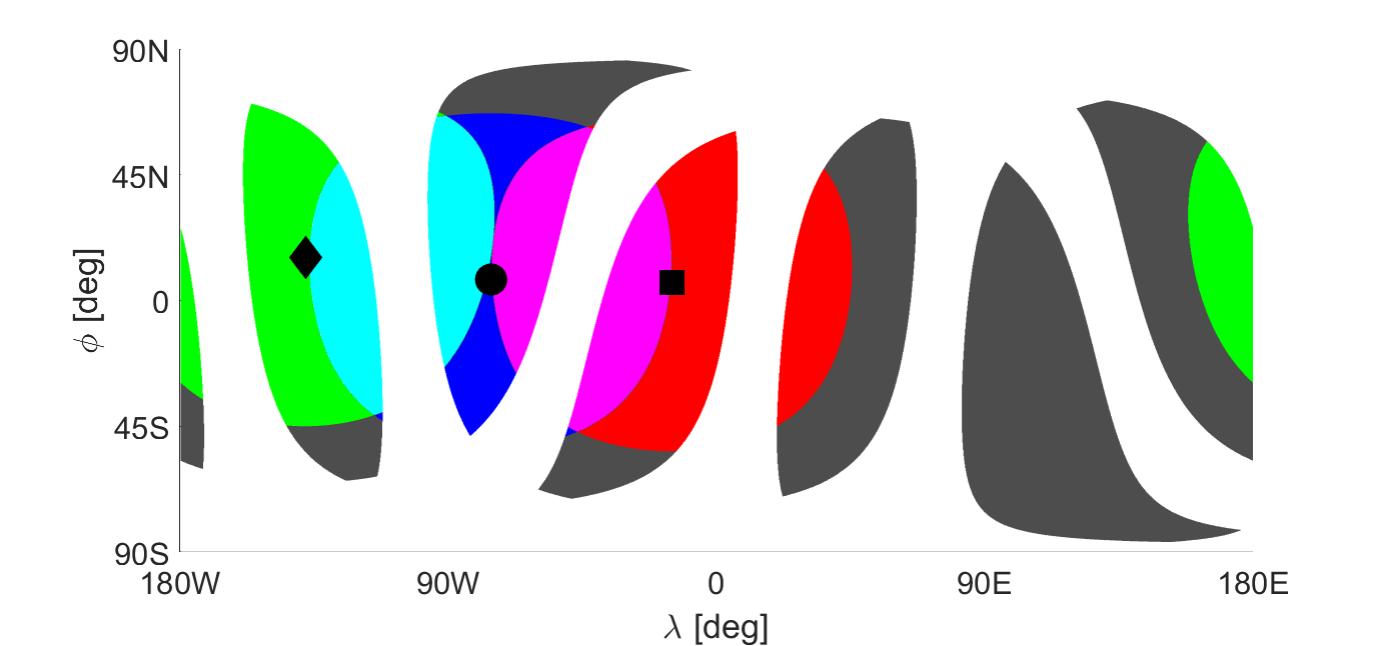}
        \caption{January}
    \end{subfigure}\\
    \begin{subfigure}[b]{.7\textwidth}
        \centering
        \includegraphics[page=6,width=.99\textwidth]{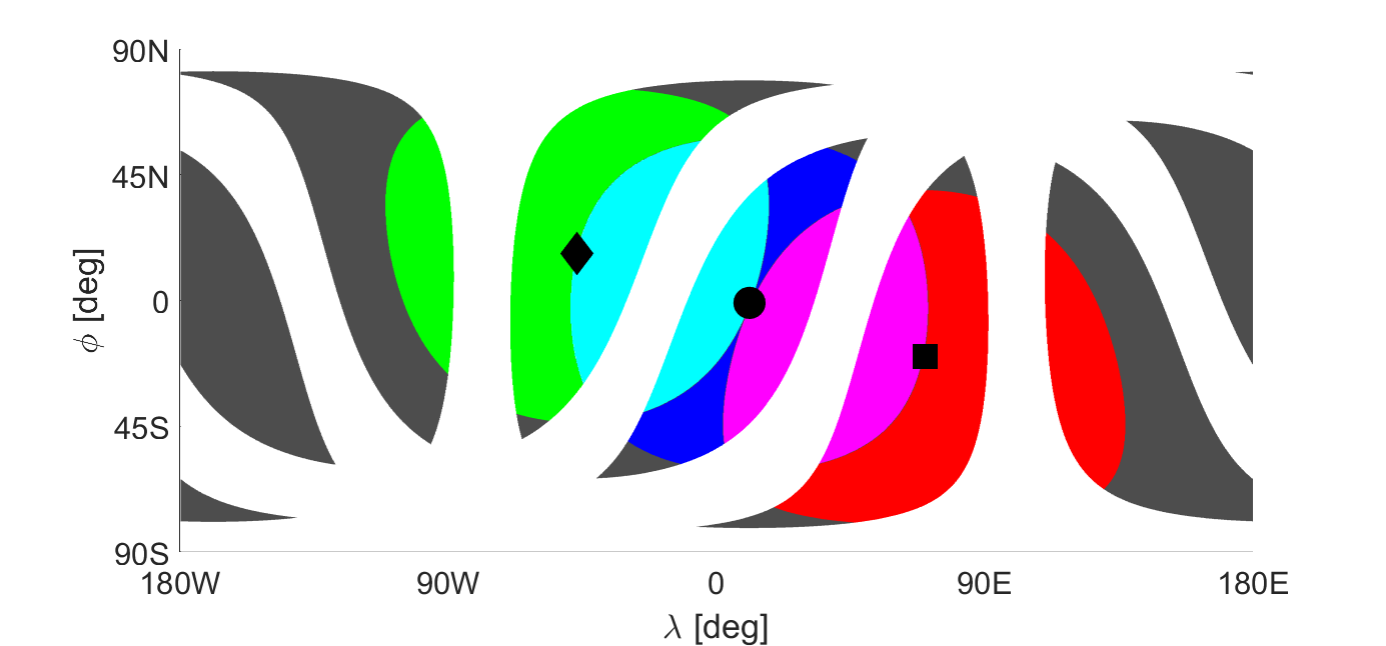}
        \caption{April}
    \end{subfigure}\\
    \begin{subfigure}[b]{.7\textwidth}
        \centering
        \includegraphics[page=7,width=.99\textwidth]{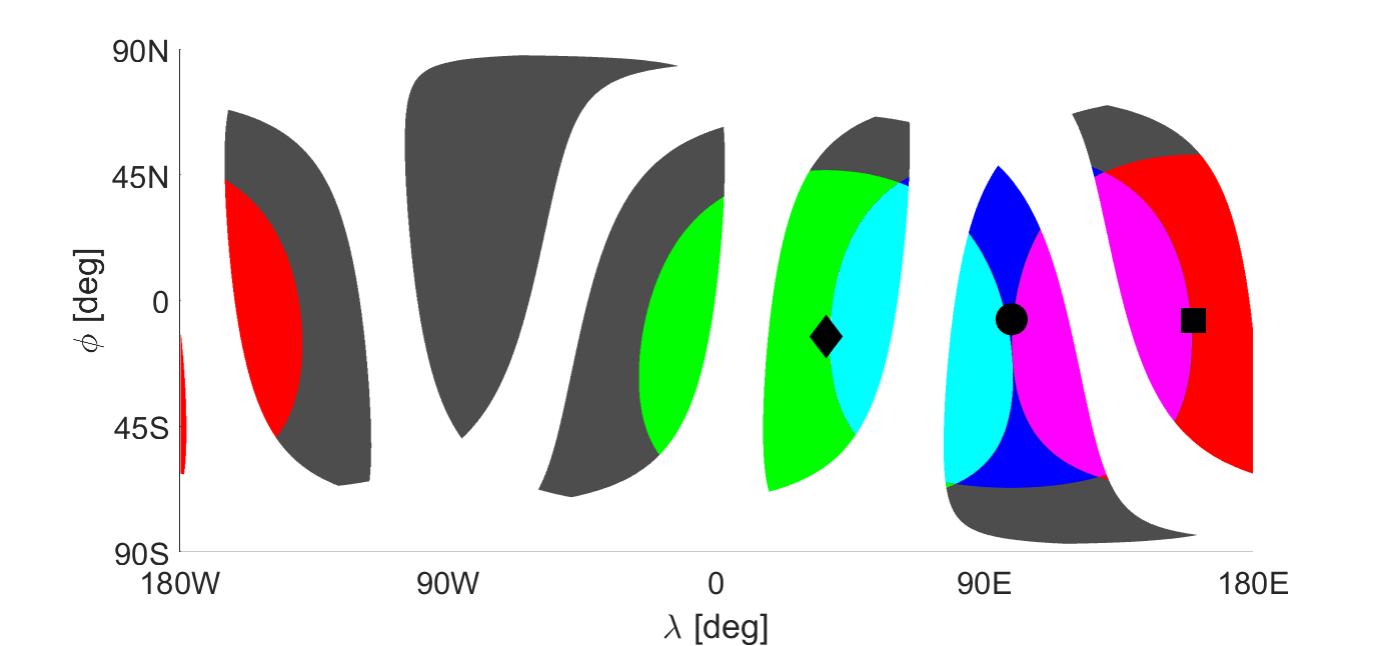}
        \caption{July}
    \end{subfigure}\\
    \begin{subfigure}[b]{.7\textwidth}
        \centering
        \includegraphics[page=8,width=.99\textwidth]{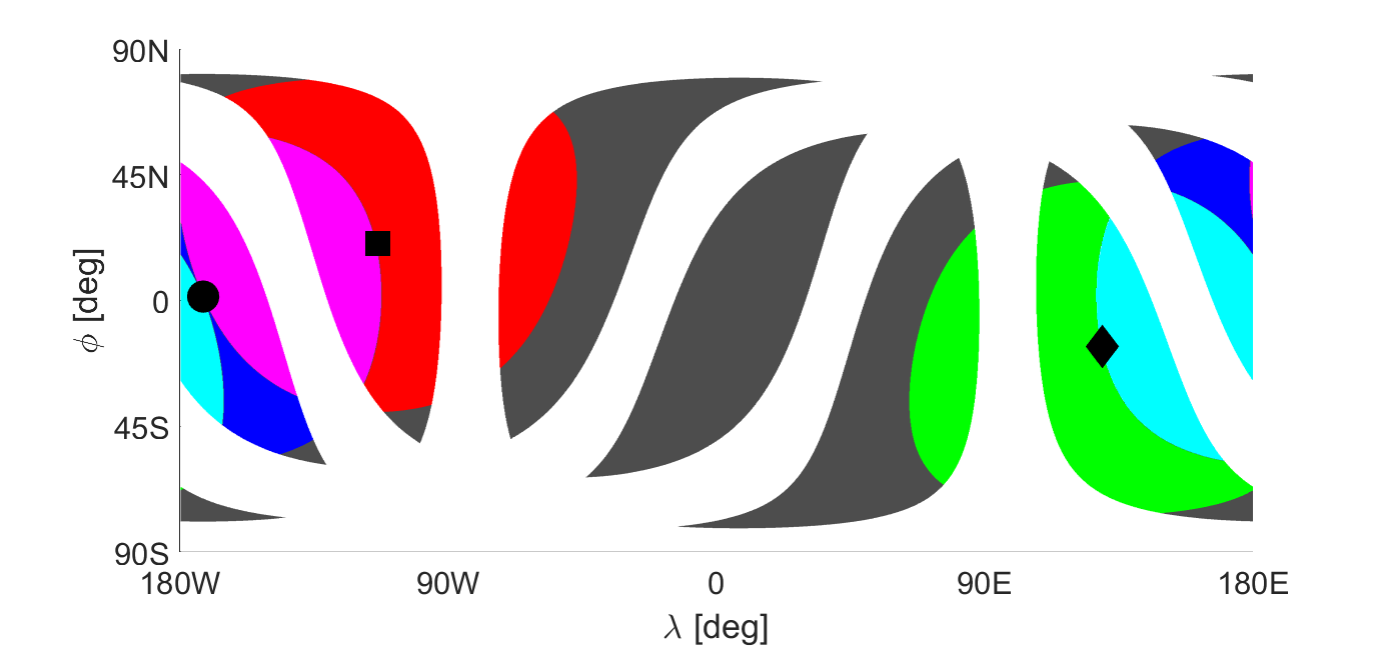}
        \caption{November}
    \end{subfigure}\\
  \end{tabular}
  \caption{Solar surface coverage with spacecraft located in planar L1 and vertical L4 and L5 periodic orbits (14.5 inclination) with $\Theta_\texttt{max}=60^\circ$.}
  \label{fig:coveragemap_145}
\end{figure*}

The figures in the Appendix section illustrate the visibility condition of the solar surface with multi-spacecraft located in L1, L4 and L5. Figure~\ref{fig:coveragemap_0} shows the coverage map of the multi-spacecraft with zero inclination. Figure~\ref{fig:coveragemap_145} shows the identical coverage map with $14.5^\circ$ inclined periodic orbit for spacecraft located in the triangular Lagrange points. The Earth's location is projected onto the solar surface with a red dot. The projected location of the L4 and L5 on the solar surface is plotted as red square and diamond, respectively. 
The visible solar surfaces of the spacecraft are shown in blue, red and green, respectively, for L1, L4 and L5, while the invisible region is in gray. The region visible by two spacecraft is denoted in magenta for L1 and L4, while in cyan for L1 and L5. The white region represents the solar limb viewed from each of the spacecraft.
All simulation assumed $\Theta_\texttt{max}=60^\circ$ for on-disk observation, and $80.08^\circ<\Theta_\texttt{max}<99.39^\circ$ for limb observation from Sun-Earth L5, which is derived from $h_{apex} = 10,000km$.

%%%%

%%%%

\clearpage

\section*{Open Research Section}
The data and code supporting the conclusions of this research are available upon request from the corresponding author.

\acknowledgments
This work was supported by the Korea Astronomy and Space Science Institute under the R\&D program (KASI2022E90190). We thank John Lee, Jae-Hung Han, Junga Hwang and Kwangsun Ryu for their constructive comments and suggestions for the L4 mission.

\bibliography{references}

\end{document}

%% ------------------------------------------------------------------------ %%
%
%  TITLE
%
%% ------------------------------------------------------------------------ %%

%\includegraphics{agu_pubart-white_reduced.eps}

\title{Supporting Information for "Insert Title"}
%
% e.g., \title{Supporting Information for "Terrestrial ring current:
% Origin, formation, and decay $\alpha\beta\Gamma\Delta$"}
%
%DOI: 10.1002/%insert paper number here%

%% ------------------------------------------------------------------------ %%
%
%  AUTHORS AND AFFILIATIONS
%
%% ------------------------------------------------------------------------ %%

% List authors by first name or initial followed by last name and
% separated by commas. Use \affil{} to number affiliations, and
% \thanks{} for author notes.
% Additional author notes should be indicated with \thanks{} (for
% example, for current addresses).

% Example: \authors{A. B. Author\affil{1}\thanks{Current address, Antartica}, B. C. Author\affil{2,3}, and D. E.
% Author\affil{3,4}\thanks{Also funded by Monsanto.}}

\authors{=Authors=}

% \affiliation{1}{First Affiliation}
% \affiliation{2}{Second Affiliation}
% \affiliation{3}{Third Affiliation}
% \affiliation{4}{Fourth Affiliation}

\affiliation{=number=}{=Affiliation Address=}
%(repeat as many times as is necessary)

%% ------------------------------------------------------------------------ %%
%
%  BEGIN ARTICLE
%
%% ------------------------------------------------------------------------ %%

% The body of the article must start with a \begin{article} command
%
% \end{article} must follow the references section, before the figures
%  and tables.

\begin{article}

%% ------------------------------------------------------------------------ %%
%
%  TEXT
%
%% ------------------------------------------------------------------------ %%

\noindent\textbf{Contents of this file}
%%%Remove or add items as needed%%%
\begin{enumerate}
\item Text S1 to Sx
\item Figures S1 to Sx
\item Tables S1 to Sx
%if Tables are larger than 1 page, upload as separate excel file
\end{enumerate}
\noindent\textbf{Additional Supporting Information (Files uploaded separately)}
\begin{enumerate}
\item Captions for Datasets S1 to Sx
\item Captions for large Tables S1 to Sx (if larger than 1 page, upload as separate excel file)
\item Captions for Movies S1 to Sx
\item Captions for Audio S1 to Sx
\end{enumerate}

\noindent\textbf{Introduction}
%Type or paste your text here. The introduction gives a brief overview of the supporting information. You should include information %about as many of the following as possible (when appropriate):
% 1. a general overview of the kind of data files;
% 2. information about when and how the data were collected or created;
% 3. a general description of processing steps used;
% 4. any known imperfections or anomalies in the data.

%\clearpage

%Delete all unused file types below. Copy/paste for multiples of each file type as needed.
\noindent\textbf{Text S1.}
%Type or paste text here. This should be additional explanatory text, such as: extended descriptions of results, full details of models, extended lists of acknowledgements etc.  It should not be additional discussion, analysis, interpretation or critique. It should not be an additional scientific experiment or paper.
%
%Repeat for any additional Supporting Text

%%Enter Data Set, Movie, and Audio captions here
%%EXAMPLE CAPTIONS

\noindent\textbf{Data Set S1.} %Type or paste caption here.
%upload your dataset(s) to AGU's journal submission site and select "Supporting Information (SI)" as the file type. Following naming %convention: ds01.

%Repeat for any additional Supporting data sets

\noindent\textbf{Movie S1.} %Type or paste caption here.
%upload your movie(s) to AGU's journal submission site and select, "Supporting Information %(SI)" as the file type. Following naming convention: ms01.

%Repeat any additional Supporting movies

\noindent\textbf{Audio S1.} %Type or paste caption here.
\end{article}
\clearpage